\makeatletter
\newcommand{\dontusepackage}[2][]{%
  \@namedef{ver@#2.sty}{9999/12/31}%
  \@namedef{opt@#2.sty}{#1}}
\makeatother
\dontusepackage{subfigure}

\documentclass[endfloat,manuscript]{geophysics}
\usepackage{lmodern}
\usepackage{amssymb,amsmath}
\usepackage{ifxetex,ifluatex}
\usepackage[usenames,dvipsnames]{color}
\usepackage{fixltx2e} % provides \textsubscript
\ifnum 0\ifxetex 1\fi\ifluatex 1\fi=0 % if pdftex
  \usepackage[T1]{fontenc}
  \usepackage[utf8]{inputenc}
\else % if luatex or xelatex
  \ifxetex
    \usepackage{mathspec}
    \usepackage{xltxtra,xunicode}
  \else
    \usepackage{fontspec}
  \fi
  \defaultfontfeatures{Mapping=tex-text,Scale=MatchLowercase}
  
\fi
\IfFileExists{upquote.sty}{\usepackage{upquote}}{}
\IfFileExists{microtype.sty}{%
\usepackage{microtype}
\UseMicrotypeSet[protrusion]{basicmath} % disable protrusion for tt fonts
}{}
\usepackage[margin=1.0in,bottom=1.5in]{geometry}
\usepackage[]{natbib}
\usepackage{listings}
\usepackage{graphicx}
\makeatletter
\def\maxwidth{\ifdim\Gin@nat@width>\linewidth\linewidth\else\Gin@nat@width\fi}
\def\maxheight{\ifdim\Gin@nat@height>\textheight\textheight\else\Gin@nat@height\fi}
\makeatother
\setkeys{Gin}{width=\maxwidth,height=\maxheight,keepaspectratio}
\usepackage{caption}
\usepackage{float}
\usepackage{algorithmic}
\usepackage{algorithm} 
\usepackage{subfig}
\ifxetex
  \usepackage[setpagesize=false, % page size defined by xetex
              unicode=false, % unicode breaks when used with xetex
              xetex]{hyperref}
\else
  \usepackage[unicode=true]{hyperref}
\fi
\hypersetup{breaklinks=true,
            bookmarks=true,
            pdfauthor={},
            pdftitle={GPU Accelerated 3D P-wave Source Free Adaptive Wavefield Reconstruction Inversion with an application to experimental VSP physical modeling data},
            colorlinks=true,
            citecolor=black,
            urlcolor=blue,
            linkcolor=black,
            pdfborder={0 0 0}}
\newcommand{\mvec}{\mathbf{m}}

\newcommand{\uvec}{\mathbf{u}}

\newcommand{\dvec}{\mathbf{d}}

\newcommand{\xvec}{\mathbf{x}}

\newcommand{\wvec}{\mathbf{w}}
\newcommand{\qvec}{\mathbf{q}}
\newcommand{\evec}{\mathbf{e}}
\newcommand{\bvec}{\mathbf{b}}
\newcommand{\pvec}{\mathbf{p}}

\newcommand{\svec}{\mathbf{s}}

\newcommand{\nsrc}{n_{\text{s}}}
\newcommand{\nrec}{n_{\text{r}}}

\newcommand{\niter}{n_{\text{it}}}
\newcommand{\ngrid}{n_{\text{g}}}

\newcommand{\nfreq}{n_{\text{f}}}

\newcommand{\Mmat}{\mathbf{M}}
\newcommand{\Amat}{\mathbf{A}}

\newcommand{\A}{\mathbf{A}}
\renewcommand{\d}{\mathbf{d}}
\newcommand{\g}{\mathbf{g}}
\renewcommand{\P}{\mathbf{P}}

\newcommand{\CC}{\mathbb{C}}

\newcommand{\RR}{\mathbb{R}}

\newcommand{\Smat}{\mathbf{S}}

\newcommand{\ffull}{f_\text{0}(\mvec,\uvec,\alpha)}

\newcommand{\fred}{f_\text{1}(\mvec,\alpha)}

\def\argmin{\mathop{\rm arg\,min}}

\DeclareMathOperator*{\minimize}{min}

\righthead{GPU Accelerated 3D SF-AWRI}
\title{GPU Accelerated 3D P-wave Source Free Adaptive Wavefield Reconstruction Inversion with an application to experimental VSP physical modeling data}
\author{  
\\\  }
\author{Zhilong Fang\textsuperscript{1}, Jingjing Zong\textsuperscript{1}
\\\textsuperscript{1}School of resources and environment, University of Electronic Science and Technology of China, Chengdu, China, \emph{611731}
\\ Corresponding author: Jingjing Zong, \text{E-mail: jjingzong@gmail.com}
}
\date{}

\begin{document}
\maketitle

\thispagestyle{empty}

\begin{abstract}
Wavefield reconstruction inversion (WRI) has been consider as a potential solution to the issue of local minima inherent in conventional full waveform inversion (FWI) methods. However, most current WRI research has been confined to 2D problems due to the computational challenges posed by solving augmented systems for optimal data-fitting wavefields. This constraint limits WRI’s applicability to realistic 3D scenarios. This study introduces a GPU-accelerated 3D source-free adaptive WRI (GPU-SF-AWRI) method that overcomes these computational barriers by adaptively controlling the computational accuracy of wavefield simulation and optimizing GPU utilization, thus enhancing its suitability for 3D applications. The inclusion of an on-the-fly source estimation technique further boosts its performance on realistic problems. Numerical experiments reveal that the proposed GPU-accelerated method achieves a 195-fold speedup compared to CPU-based approaches. By incorporating adaptive accuracy and total variation regularization, we attain a 2-fold speedup while maintaining inversion accuracy. We applied the GPU-SF-AWRI method to numerical and actual Vertical Seismic Profiling (VSP) physical modeling P-wave data, confirming its efficacy in addressing real data challenges and mitigating local minima associated with conventional FWI.

\end{abstract}

\section{Introduction}\label{introduction}
Full Waveform Inversion (FWI) has garnered considerable attention as a promising tool for automatically generating high-resolution subsurface images in both the applied geophysics industry and seismology academia \citep{Tarantola1982FWI, Pratt98, VirieuxOverview2009, tromp2019seismic}. Despite its successful applications in the practical seismic exploration, conventional FWI still struggles with a significant challenge known as local minima, commonly referred to as 'cycle-skipping'. This issue arises when the initial velocity model is inaccurate and low-frequency data are unavailable \citep{warner2013full}. Consequently, to ensure the success of FWI applications, the conventional approach typically relies on a precise initial model or data containing sufficient low-frequency components.

To tackle the challenge of cycle skipping, researchers have explored alternative misfit functions beyond the conventional least-squared loss function. These approaches encompass correlation-based misfit functions \citep{van2010correlation}, envelope-based misfit functions \citep{wu2013ultra, chi2014full}, Weiner filter-based misfit functions \citep{warner2016adaptive}, and Wasserstein-based misfit functions \citep{engquist2014application, metivier2016measuring, yang2018application}. Concurrently, others have addressed cycle skipping by broadening the search space \citep{barnier2023full, operto_extending_2023}, proposing methods such as wavefield reconstruction inversion \citep{vanleeuwen2015IPpmp, fang2018source}, extended sources waveform inversion \citep{huang2017full}, lift and relax waveform inversion \citep{fang2020lift}, and wave-equation migration velocity analysis (WENVA) \citep{symes2008migration, li2014wave, yang2020least}.

WRI, recognized as a promising solution for addressing cycle skipping, has received considerable interest among researchers \citep{operto_extending_2023}. Its applications extend to seismic exploration \citep{vanLeeuwen2013GJImlm}, microseismic imaging \citep{aghamiry2022admm}, and uncertainty quantification for subsurface structures \citep{fang2018uncertainty}. Moreover, researchers have extended the original frequency domain acoustic WRI to include time domain WRI \citep{rizzuti2021dual} and elastic WRI \citep{song2023elastic}. Despite these advancements, the focus on 2D problems persists in WRI-related research due to the computational challenges of solving large data-augmented systems. To alleviate computational costs, researchers have attempted to formulate dual problems of WRI, replacing the direct solution of the data-augmented system with the solution of multiple partial differential equations (PDEs) . This approach yields approximate solutions to the original data-augmented system with computational costs equivalent to solving several PDEs \citep{rizzuti2021dual, gholami2022extended, lin2023fast}. However, despite these efforts, the substantial computational cost has continued to constrain the exploration of 3D WRI, limiting its application to real-world problems.

In this investigation, we aim to expand the application of WRI from two-dimensional (2D) contexts to three-dimensional (3D) scenarios. As previously noted, the primary computational challenge involves solving the large, sparse, linear data augmented system. For a system with \(\ngrid\) grid points, the computational cost of direct solvers is proportional to \(\mathcal{O}(\ngrid^{3/2})\). In contrast, sparse iterative solvers like Least Squares QR method (LSQR) \citep{paige1982lsqr} have a computational cost of \(\mathcal{O}(\ngrid)\) and require significantly less storage than direct solvers, making them more suitable for 3D problems. The overall computational cost of iterative solvers depends on the number of iterations \(\niter\) and the time required for each iteration. To address these challenges, we propose an adaptive approach that reduces the number of iterations required and leverages GPU acceleration to enhance the computational efficiency of each iteration.

In conventional WRI, the data-augmented system is solved with high accuracy to obtain a precise gradient for updating the model. However, such a high accuracy may not be necessary for generating a descent direction, particularly in the early stages of inversion. We propose an adaptive WRI (AWRI) approach that dynamically controls accuracy throughout the inversion process. In the early stages, AWRI employs lower accuracy to accelerate model updates. As the inversion progresses, AWRI adaptively increases the accuracy based on the optimization status. In the later stages, higher accuracy is used to ensure precise solutions. This adaptive accuracy strategy is similar to the method introduced by \citet{van20143d} for  conventional 3D FWI . The trade-off for using lower accuracy is the introduction of artifacts in the model, which can necessitate a rapid increase in accuracy. To mitigate these artifacts, we incorporate total variation (TV) regularization. TV regularization has been demonstrated to be effective in suppressing artifacts and preserving edges in conventional FWI \citep{esser2016tvr, aghamiry2019admm, fang2024source}. The introduction of TV regularization allows us to sustain the lower accuracy for extended periods, thereby increasing computational speed while preserving inversion accuracy.

In addition to accelerating WRI through adaptive strategies, we aim to expedite computations within each iteration by leveraging Graphics Processing Units (GPUs). The main computational expense in each iteration of LSQR arises from matrix-vector multiplications, which involve straightforward arithmetic operations such as multiplication and addition. Furthermore, these operations are inherently parallelizable, as the products of matrix rows and vectors are independent across different rows. These characteristics make matrix-vector multiplication well-suited for GPUs, allowing them to harness thousands of computational units effectively. Therefore, we have implemented the iterative LSQR method on GPUs to accelerate computations in our proposed 3D AWRI framework.

We first conduct numerical experiments to assess the acceleration performance of the proposed GPU framework. To confirm the practicality of the 3D GPU-Accelerated AWRI in real-world scenarios, we also apply it to the analysis of experimental Vertical Seismic Profiling (VSP) physical modeling data. By employing a down-scaled model and survey configuration and, subsequently, ultrasonic source-receiver pairs, the physical modeling experiment provides actual wave propagation which serves as an analog for field-scale seismic acquisition. Compared to surface seismic data, the VSP wavefield provides a clearer depiction of elastic wave propagation and conversion with depth \citep{stewart2002converted}, offering high-quality wavefield data for WRI approaches. Additionally, due to the asymmetry and uneven illumination in the VSP acquisition system, it is more sensitive to velocity errors compared to surface seismic geometry, resulting in more rigorous requirements for the accuracy of the initial velocity model \citep{blias20153d}. Therefore, the VSP survey geometry is ideal for validating the proposed 3D GPU-AWRI method. This experiment represents the first application of WRI to real data, highlighting the novelty and reliability of the proposed approach. The real-data problem presents additional challenges due to the lack of a highly accurate source function, complicating the inversion process. To address this, we extend the on-the-fly source estimation method from 2D WRI \citep{fang2018source} to our 3D framework, resulting in a source-free AWRI approach (SF-AWRI). 

The results from the numerical experiments indicate that the GPU framework achieves an acceleration factor of 195$\times$ compared to the CPU-based framework. In a VSP scenario, the numerical example demonstrates the proposed GPU-SF-AWRI's effectiveness in mitigating local minima. Additionally, applying the method to real experimental VSP physical modeling data showcases its feasibility for inversions with authentic datasets. To our knowledge, this represents the first real-data VSP application of a WRI-related method. Remarkably, the method yields satisfactory inversion results without requiring a precise source function.

\section{Methodology}\label{methodology}

\subsection{3D Source free wavefield reconstruction inversion}\label{SFWRI}

Consider a seismic inversion problem that uses a dataset $\mathbf{d} \in \CC^{\nsrc\times\nfreq\times\nrec}$ containing $\nsrc$ sources, $\nrec$ receivers, and $\nfreq$ frequencies. The conventional FWI tries to reconstruct the unknown model parameter  $\mvec\in \RR^{\ngrid}$ (squared slowness) by solving the following PDE-constrained optimization problem: 
% In the process of FWI, we address the following least-squares problem for the discretized $\ngrid$-dimensional unknown medium parameters, denoted by $\mvec \in \RR^{\ngrid}$: 
%
\begin{equation}
\begin{split}
   \minimize_{\uvec,\,\mvec,\, \alpha}\ffull&=\frac{1}{2}\sum_{i=1}^{\nsrc}\sum_{j=1}^{\nfreq}\|\P_{i}\uvec_{i,j}-{\d}_{i,j}\|^{2}_{2},  \\
   \hspace{8mm} \text{ subject to } &\A_{j}(\mvec)\uvec_{i,j} = \alpha_{i,j}\evec_{i}.  
\end{split}
\label{eq:FWI}
\end{equation}
% %
In this optimization problem, besides the unknown parameter $\mvec$, we also need to optimize over the unknown pressure wavefield $\uvec \in \CC^{\nsrc\times\nfreq\times\ngrid}$ and the unknown source function $\alpha$.  The indices $i$ and $j$ correspond to the source and frequency, respectively. The matrix $\A_{j}(\mvec) = \omega_{j}^{2}\text{diag}(\mvec) + \Delta$ symbolizes the discretized 3D Helmholtz matrix at angular frequency $\omega_{j}$. Here, $\Delta = \frac{\partial^2}{\partial x^2}+\frac{\partial^2}{\partial y^2}+\frac{\partial^2}{\partial z^2}$ represents the discretized 3D Laplacian operator, and the matrix $\P_{i}$ acts as a restriction operator for the receiver locations. Additionally,  $\evec_{i}$ is a unit vector representing the location of the $i^{\text{th}}$ source. For simplicity, we will omit the explicit dependence of $\A_j(\mvec)$ on the model parameter $\mvec$ in the subsequent notation.

The PDE-constrained optimization problem presented in equation \ref{eq:FWI} requires the inversion of parameters within an exceptionally large space $\RR^{\ngrid} \times \CC^{\nsrc\times\nfreq\times(\ngrid+1)}$. This size proves impractical for current hardware, including the most extensive High-Performance Computing Cluster, which faces limitations in storing the vast array of unknown parameters and executing the inversion process. To surmount the challenge posed by the extensive storage requirements, a widely recognized approach is the reduced method \citep{VirieuxOverview2009}, in which the PDE constraint and the unknown wavefields are eliminated by explicitly solving the PDE as follows
\begin{equation}
    \uvec_{i,j} = \alpha_{i,j}\Amat_{j}^{-1}\evec_{i}.
\label{eq:PDEsolve}
\end{equation}
Substituting the wavefield $\uvec$ in equation \ref{eq:FWI} and removing the PDE constraint, we can obtain the following objective function concerning the model parameter $\mvec$ and source function $\alpha$:
\begin{equation}
\minimize_{\mvec,\alpha}\fred=\frac{1}{2}\sum_{i=1}^{\nsrc}\sum_{j=1}^{\nfreq}\|\alpha_{i,j}\P_{i}\A_{j}^{-1}\evec_{i}-{\d}_{i,j}\|^{2}_{2}.
\label{eq:FWI2}
\end{equation}
The optimization described in equation \ref{eq:FWI2} focuses on searching for unknown parameters in the space of $\RR^{\ngrid}\times\CC^{\nsrc\times\nfreq}$, which is considerably smaller than the space in equation \ref{eq:FWI}. The substantial reduction in the dimensionality of the search space allows researchers and engineers to conduct FWI on existing hardware. However, this dimensionality reduction comes with the trade-off of increased nonlinearity in the objective function $\fred$ concerning $\mvec$. This is because the correlation of the objective function $\fred$ with $\mvec$ operates through the nonlinear operator $\A_{j}^{-1}\qvec_{i,j}$ rather than the linear operator $\A_{j}\uvec_{i,j}$. Consequently, the reduced objective function $\fred$ may contain local minima \citep{warner2013full}, presenting greater challenges in determining the optimal medium parameters using only local derivative information.

To address the local minima issue in the reduced FWI method in equation \ref{eq:FWI2}, \citet{fang2018source} extends WRI \citep{vanLeeuwen2013GJImlm, vanleeuwen2015IPpmp} to scenario with unknown source function, and propose the following source independent WRI objective function: 

\begin{equation}
  \minimize_{\mvec,\, \uvec,\, \alpha} f_{2}(\mvec,\,\uvec,\,\alpha)=\frac{1}{2}\sum_{i=1}^{\nsrc}\sum_{j=1}^{\nfreq}\|\P_{i}\uvec_{i,j}-{\d}_{i,j}\|^{2}_{2} + \lambda^{2}\|\A_{j}\uvec_{i,j}-\alpha_{i,j}\mathbf{e}_{i,j}\|^{2}_{2}.
\label{eq:ObjPen}
\end{equation}

In this optimization problem, we treat the wavefields $\uvec$ as unknown variables, departing from the elimination process in FWI. Instead, we relax the PDE constraint by adding an $\ell_2$-norm penalty term in the objective function. The scalar parameter $\lambda$ governs the trade-off between the data and PDE misfits. As $\lambda$ increases, the wavefield becomes more strongly constrained by the wave equation. Consequently, the objective function $f_{2}(\mvec,\,\uvec,\,\alpha)$ in equation~\ref{eq:ObjPen} converges towards the objective function of the reduced objective function in equation~\ref{eq:FWI2} \citep{vanleeuwen2015IPpmp}. Unlike the nonlinear objective function in equation~\ref{eq:FWI2}, the WRI objective $f_{2}(\mvec,\,\uvec,\,\alpha)$ exhibits a bi-quadratic nature with respect to $\mvec$ and the augmented vector $\xvec = \begin{bmatrix} \uvec \\ \alpha \end{bmatrix}$. Indeed, we can write the following compact form for the objective $f_{3}(\mvec,\,\xvec) = f_{2}(\mvec,\,\uvec,\,\alpha)$:

\begin{equation}
    f_{3}(\mvec,\,\xvec) = \frac{1}{2}\sum_{i=1}^{\nsrc}\sum_{j=1}^{\nfreq}\|\Smat_{i,j}\xvec_{i,j}-\bvec_{i,j}\|^2_2,
\label{eq:CmpctWRIobj}
\end{equation}
where
\begin{equation}
\Smat_{i,j} = \begin{bmatrix}
  \lambda\A_{j} & -\lambda\mathbf{e}_{i,j} \\
  \P_{i} & 0
  \end{bmatrix}, \quad \text{and} \quad \bvec_{i,j} = \begin{bmatrix} 0 \\ \dvec_{i,j} \end{bmatrix}
\label{eq:Sandb}
\end{equation}
Clearly, for a fixed $\mvec$, the objective $f_{3}(\mvec,\,\xvec) = f_{2}(\mvec,\,\uvec,\,\alpha)$ is quadratic with respect to $\xvec$, and vice versa. Additionally, \citet{vanLeeuwen2013GJImlm} and \citet{vanleeuwen2015IPpmp} have highlighted that WRI explores a broader search space by not strictly adhering to the PDE constraints, making WRI potentially less susceptible to numerous local minima compared to the conventional FWI.

The optimization problem described in equation~\ref{eq:ObjPen} entails minimizing the wavefields $\uvec$, medium parameters $\mvec$, and source function $\alpha$. Conventional optimization methods such as gradient descent and limited-memory Broyden–Fletcher–Goldfarb–Shanno (l-BFGS) methods \citep{Nocedal:2000} necessitate storing these three unknown variables. However, the memory cost associated with storing $\uvec$ can be exceptionally high because $\uvec \in \CC^{\ngrid\times\nfreq\times\nsrc}$. To tackle this storage challenge, inspired by the source estimation method for 2D WRI \citep{fang2018source}, we can leverage the aforementioned bi-quadratic behavior of $f_{3}(\mvec,\xvec)$ concerning $\mvec$ and $\xvec$ and use the variable projection method \citep{golub2003separable} to propose a 3D source-free WRI method. Considering the bi-quadratic property, for a fixed $\mvec$, there is an analytical solution $\overline{\xvec} = \begin{bmatrix} \overline{\uvec} \\ \overline{\alpha} \end{bmatrix}$ to minimize $f_{3}(\mvec,\xvec)$:

\begin{equation}
\begin{aligned}
\overline{\xvec}=\begin{bmatrix}
  \overline{\uvec} \\
  \overline{\alpha} 
\end{bmatrix} 
&= \argmin_{\uvec,\alpha} f_2(\mvec,\uvec,\alpha), \\
\text{with} \quad \overline{\xvec}_{i,j} = \begin{bmatrix} \overline{\uvec}_{i,j}(\mvec) \\
  \overline{\alpha}_{i,j}(\mvec) \end{bmatrix} &= \begin{bmatrix}
  \lambda^{2}{\A_{j}^{\top}\A_{j}}+\P_{i}^{\top}\P_{i} & -\lambda^{2}\A^{\top}_{j}{\mathbf{e}_{i,j}} \\
  -\lambda^{2}\mathbf{e}_{i,j}^{\top}\A_{j} & \lambda^{2}\mathbf{e}_{i,j}^{\top}\mathbf{e}_{i,j}
  \end{bmatrix}^{-1} \begin{bmatrix} \P_{i}\d_{i,j} \\ \mathbf{0}\end{bmatrix}.
\end{aligned}
\label{eq:SimpObj}
\end{equation}

Substituting $\begin{bmatrix}
  \overline{\uvec} \\
  \overline{\alpha} 
\end{bmatrix}$ for $\begin{bmatrix}
  \uvec \\
  \alpha
  \end{bmatrix}$  in equation \ref{eq:ObjPen}, we can  obtain an objective function that only depends on $\mvec$:

\begin{equation}
   f_4(\mvec) = f_3\big(\mvec,\overline{\xvec}(\mvec)\big) = f_2\big (\mvec,\overline{\uvec}(\mvec),\overline{\alpha}(\mvec)\big ).
\label{eq:VarObj2}
\end{equation}
We can use the chain-rule to compute the gradient $\g_4(\mvec)$ of $f_4(\mvec)$ as follows:
%
%\begin{equation}
%\begin{aligned}
%  \g_4(\mvec) = &\nabla_{\mvec}f_{4}(\mvec) = \nabla_{\mvec}f_3(\mvec,\xvec)\rvert_{\xvec=\overline{\xvec}(\mvec)}\\
%  &+\nabla_{\xvec}f_3(\mvec,\xvec)\rvert_{\xvec=\overline{\xvec}(\mvec)}\nabla_{\mvec}\xvec \\
%\end{aligned}
%\label{eq:VarG2}
%\end{equation}
%
%
\begin{equation}
\begin{aligned}
  \g_4(\mvec) = &\nabla_{\mvec}f_{4}(\mvec) =\sum_{i=1}^{\nsrc}\sum_{j=1}^{\nfreq}\lambda^2\omega_{i}^{2}\text{diag}\Big (\text{conj}\big (\overline{\uvec}_{i,j}(\mvec)\big )\Big )\big (\A_{j}\overline{\uvec}_{i,j}(\mvec)-\overline{\alpha}_{i,j}(\mvec){\evec}_{i,j}\big )
\end{aligned}
\label{eq:VarG2}
\end{equation}

%Since $\overline{\xvec}$ minimizes $f_3(\mvec,\xvec)$ for given $\mvec$, we have $\nabla_{\xvec}f_3(\mvec,\xvec)\rvert_{\xvec=\overline{\xvec}(\mvec)} = 0$. Thus, we can simplify the gradient $\g_4(\mvec)$ to:

%\begin{equation}
%\begin{aligned}
%  \g_4(\mvec) &= \nabla_{\mvec}f_4(\mvec) = \nabla_{\mvec}f_3(\mvec,\xvec)\rvert_{\xvec=\overline{\xvec}(\mvec)} \\
%  &=\sum_{i=1}^{\nsrc}\sum_{j=1}^{\nfreq}\lambda^2\omega_{i}^{2}\text{diag}\Big (\text{conj}\big (\overline{\uvec}_{i,j}(\mvec)\big )%\Big )\big (\A_{j}\overline{\uvec}_{i,j}(\mvec)-\overline{\alpha}_{i,j}(\mvec){\evec}_{i,j}\big ).
%\end{aligned}
%\label{eq:VarG3}
%\end{equation}

\begin{figure}
\centering
{\includegraphics[width=0.90\hsize]{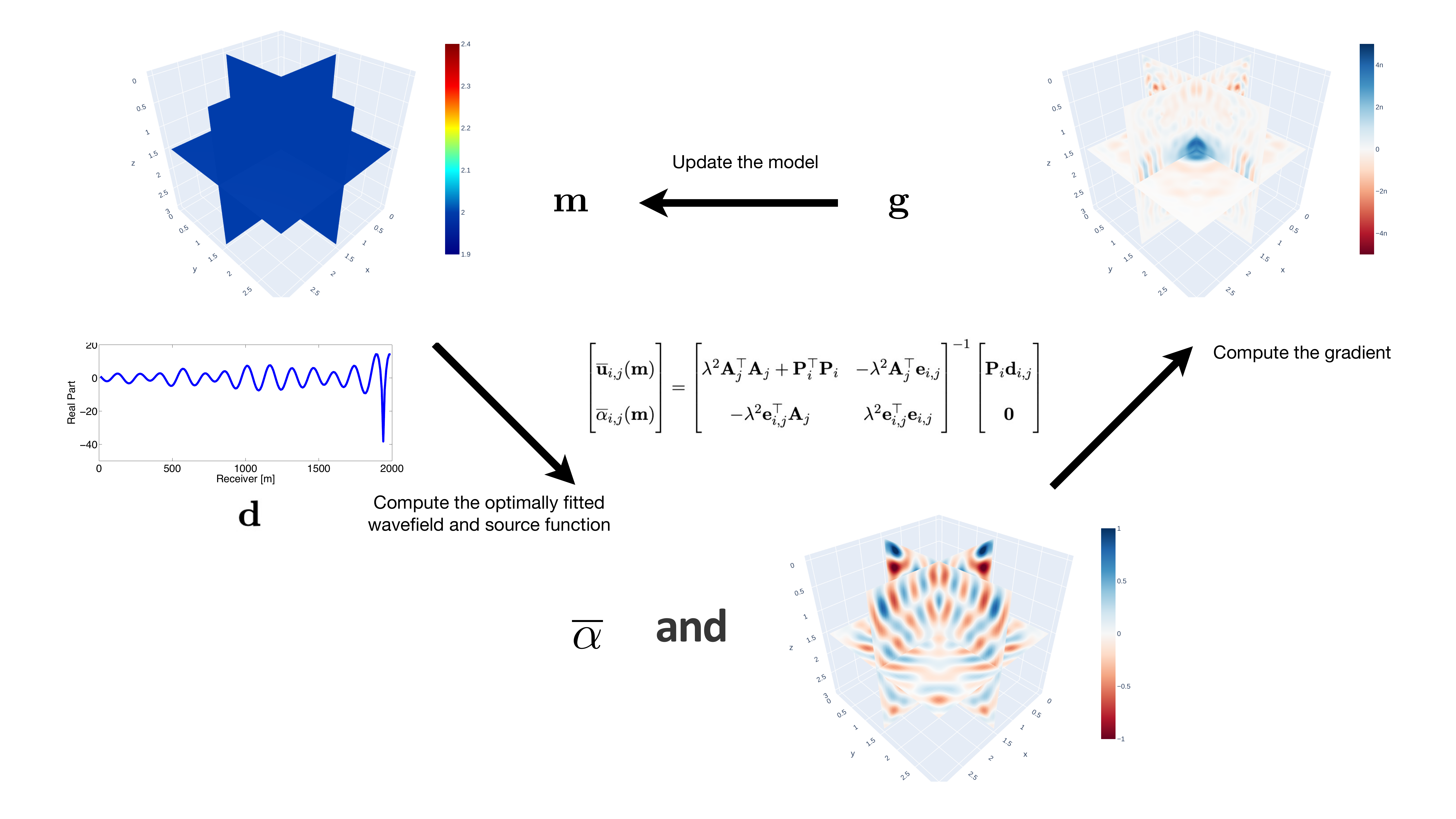}}
\caption{A 2D example to show the process of source estimation, gradient calculation, and model updating at each iteration.}
\label{fig:SI}
\end{figure}

Leveraging the gradient $\g_4(\mvec)$, we can apply local search methods such as gradient descent and the l-BFGS method for minimizing over $\mvec$. Figure~\ref{fig:SI} illustrates the process of source estimation, gradient calculation, and model updating at each iteration. As noted in equation \ref{eq:VarG2}, the computation of $\g_4(\mvec)$ does not involve additional PDE solvers or the inversion of large-scale matrices. The primary computational cost lies in inverting the data-augmented system in equation \ref{eq:SimpObj}, requiring an additional acceleration strategy.

\subsection{GPU-Accelerated 3D Source-Free Adaptive WRI}\label{GPU1}
The primary computational burden of 3D-SF-WRI lies in solving the optimization problem in equation \ref{eq:SimpObj} or equivalently inverting the matrix

\begin{equation}
    \Mmat_{i,j} = \Smat_{i,j}^{\top}\Smat_{i,j}=\begin{bmatrix}
  \lambda^{2}{\A_{j}^{\top}\A_{j}}+\P_{i}^{\top}\P_{i} & -\lambda^{2}\A^{\top}_{j}{\mathbf{e}_{i,j}} \\
  -\lambda^{2}\mathbf{e}_{i,j}^{\top}\A_{j} & \lambda^{2}\mathbf{e}_{i,j}^{\top}\mathbf{e}_{i,j}
  \end{bmatrix},
\label{Mmatrix}
\end{equation}
which can easily surpass the size of $10^9 \times 10^9$ for 3D problems. Given the variation in source locations, the vector $\evec_{i,j}$ changes accordingly for each source, resulting in distinct matrices $\Smat_{i,j}$ and $\Mmat_{i,j}$ for different sources. Consequently, sparse direct solvers with a computational complexity of $\mathcal{O}(n^{3/2})$ are impractical for inverting the matrix $\Mmat_{i,j}$. Therefore, instead of straightforwardly inverting the matrix $\Mmat_{i,j}$, we employ the iterative solver - LSQR method, as illustrated in Algorithm \ref{algo:lsqr}, to solve the linear data-fitting optimization problem in equation~\ref{eq:SimpObj}. Compared to Conjugated Gradient method, LSQR does not need explicitly form the sparse matrix $\Mmat_{i,j}$, whose storage cost is twice of that of $\Smat_{i,j}$. The computational cost of the LSQR algorithm is directly linked to both the number of iterations and the computation required in each iteration. In this work, we introduce an adaptive WRI (AWRI) method  with controlled accuracy designed to reduce the number of iterations. Additionally, we leverage GPU acceleration to enhance computational efficiency for each LSQR iteration.

\begin{algorithm}
  \caption{LSQR Algorithm to solve $\min_{\xvec} \frac{1}{2}\|\Smat\xvec - \bvec\|^2_{2}$}
  \label{algo:lsqr}
  \begin{algorithmic}[1]
    \STATE Initialize: $\xvec_0$, $\wvec_0 = \bvec - \Smat \xvec_0$, $\beta_0 = \lVert \wvec_0 \rVert_2$, $\svec_0 = \wvec_0 / \beta_0$, $\pvec_0 = \Smat^{\top}\svec_0 / \beta_0$ \\
    \STATE $\text{repeat}$ \\
    % \For{$k = 1, 2, \ldots$}
    \STATE $\qquad$ $\alpha_k = \lVert \Smat \pvec_k \rVert_2 / \lVert \pvec_k \rVert_2$ \\
    \STATE $\qquad$ $\pvec_{k+1} = \Smat^{\top} \svec_k - \alpha_k \pvec_k$ \\
    \STATE $\qquad$ $\beta_{k+1} = \lVert \pvec_{k+1} \rVert_2$
    \STATE $\qquad$ \hbox{if }{$\beta_{k+1} = 0$} \\
    \STATE $\qquad$ $\qquad$ Algorithm terminates: exact solution found \\
    \STATE $\qquad$ $\qquad$ \textbf{break} \\
    \STATE $\qquad$ \hbox{end if}\\
    \STATE $\qquad$ $\svec_{k+1} = \pvec_{k+1} / \beta_{k+1}$ \\
    \STATE $\qquad$ $\wvec_{k+1} = \Smat^{\top} \svec_{k+1}$ \\
    \STATE $\qquad$ $\gamma_k = \lVert \wvec_{k+1} \rVert_2 / \beta_{k+1}$ \\
    \STATE $\qquad$ $\xvec_{k+1} = \xvec_k + \alpha_k \svec_k + \gamma_k \svec_{k+1}$ \\
    \STATE $\text{end repeat}$ \\
    \STATE $\text{return } \mathbf{x}_{k+1} \text{ as the result}$
  \end{algorithmic}
\end{algorithm}

\subsubsection{SF-AWRI with total variation regularization}

During the inversion, we intend to solve the linear system in equation~\ref{eq:SimpObj} with a high accuracy to ensure an accurate direction to update the model. However, in the early stage of the inversion, such a high accuracy may not be necessary \citep{van20143d}. A wavefield with a relative low accuracy may also provide a decent direction. As a result, we can dynamically adjust the accuracy for the solution of the linear system, yielding an adaptive WRI. 

Reducing accuracy generally diminishes the quality of the inverted model. To preserve inversion quality while taking advantage of enhanced computational speed, we propose integrating TV regularization into the process. TV regularization has proven effective in the conventional FWI for edge preservation and artifact suppression \citep{esser2016tvr}. In our approach, TV regularization counteracts the artifacts introduced by reduced accuracy, enabling us to achieve both accelerated inversion and high-quality results. By regularizing the model parameters with TV, AWRI can maintain low accuracy for extended periods, further accelerating the inversion process.We ultimately aim to solve the following TV-regularized problem:
\begin{equation}
    f_{4}(\mvec,\,\xvec) = \frac{1}{2}\sum_{i=1}^{\nsrc}\sum_{j=1}^{\nfreq}\|\Smat_{i,j}\xvec_{i,j}-\bvec_{i,j}\|^2_2+\gamma\|\mvec\|_{TV},
\label{eq:TVAWRI}
\end{equation}
where $\|\cdot\|_{TV}$ denotes the TV norm of a vector and $\gamma$ is the weighting parameter to balance the misfit and TV norm.

Algorithm \ref{algo:AWRI} outlines the pseudocode for the proposed SF-AWRI method incorporating TV regularization. The inversion process begins with a relatively high tolerance \(\epsilon\) for LSQR. In the \(k^{\text{th}}\) iteration, we first compute the objective function \(f_{4}^{(k)}\), the gradient \(\mathbf{g}_{4}^{(k)}\), and the l-BFGS Hessian \(\mathbf{H}_{4}^{(k)}\) using the current model \(\mathbf{m}^{(k)}\) and the current tolerance \(\epsilon\). Next, we update the model parameters to obtain a temporary model \(\tilde{\mathbf{m}}^{(k)}\) and compute the corresponding objective function \(\tilde{f}_{4}^{(k)}\). If the new objective function \(\tilde{f}_{4}^{(k)}\) is smaller than the previous one \(f_{4}^{(k)}\), we set $\mvec^{(k+1)}=\tilde{\mvec}^{(k)}$ and proceed to the next iteration. If not, we reduce the tolerance by setting \(\epsilon = 0.5\epsilon\) and attempt the model parameter update again. This algorithm achieves rapid convergence in the early stages of inversion while maintaining precision in the later stages.

\begin{algorithm}
  \caption{Source Free AWRI with TV regularization}
  \label{algo:AWRI}
  \begin{algorithmic}[1]
    \STATE Initialize: $\mvec_0, \lambda, \gamma, \epsilon_{0}$ \\
    \STATE $\epsilon = \epsilon_{0}$
    \STATE $\text{repeat}$ \\
    % \For{$k = 1, 2, \ldots$}
    \STATE $\qquad$ compute $\overline{\xvec}_{k}$ by equation~\ref{eq:SimpObj} with tolerance $\epsilon$ \\
    \STATE $\qquad$ compute the objective function  $f_{4}^{(k)}$, gradient $ \mathbf{g}_{4}^{(k)}$, and l-BFGS Hessian $\mathbf{H}_{4}^{(k)}$\\
    \STATE $\qquad$ update the model by $\tilde{\mvec}^{(k)} = \mvec^{(k)}- \mathbf{H}^{(k)}_{4}\mathbf{g}_{4}^{(k)}$\\
    \STATE $\qquad$ compute $\tilde{\xvec}_{k}$ with $\tilde{\mvec}^{(k)}$ by equation~\ref{eq:SimpObj} with tolerance $\epsilon$ \\
    \STATE $\qquad$ compute the objective function  $\tilde{f}_{4}^{(k)}$\\
    \STATE $\qquad$ if $\tilde{f}_{4}^{(k)} > {f}_{4}^{(k)}$ \\
    \STATE $\qquad$ $\qquad$ $\epsilon = 0.5\epsilon$\\
    \STATE $\qquad$ $\qquad$ go back to line 4\\
    \STATE $\qquad$ else\\
    \STATE $\qquad$ $\qquad$ $\mvec^{(k+1)}=\tilde{\mvec}^{(k)}$\\
    \STATE $\qquad$ \hbox{end if}\\
    \STATE $\text{end repeat}$ \\
    \STATE $\text{return } \mathbf{m}^{(k+1)} \text{ as the result}$
  \end{algorithmic}
\end{algorithm}

% \begin{algorithm}
% \caption{Conjugate Gradient Algorithm to solve $\Mmat \xvec = \mathbf{b}$}
% \label{algo:cg}
% \begin{algorithmic}[1]
% \STATE $\mathbf{r}_0 := \mathbf{b} - \mathbf{M x}_0$
% \STATE  $\hbox{if } \mathbf{r}_{0} \text{ is sufficiently small, then return } \mathbf{x}_{0} \text{ as the result}$\\
% \STATE $\mathbf{p}_0 := \mathbf{r}_0$ \\
% \STATE $k := 0$ \\
% \STATE $\text{repeat}$ \\
% \STATE $\qquad \alpha_k := \frac{\mathbf{r}_k^\mathsf{T} \mathbf{r}_k}{\mathbf{p}_k^\mathsf{T} \mathbf{M p}_k}$  \\
% \STATE $\qquad \mathbf{x}_{k+1} := \mathbf{x}_k + \alpha_k \mathbf{p}_k$ \\
% \STATE $\qquad \mathbf{r}_{k+1} := \mathbf{r}_k - \alpha_k \mathbf{M p}_k$ \\
% \STATE $\qquad \hbox{if } \mathbf{r}_{k+1} \text{ is sufficiently small, then exit loop}$ \\
% \STATE $\qquad \beta_k := \frac{\mathbf{r}_{k+1}^\mathsf{T} \mathbf{r}_{k+1}}{\mathbf{r}_k^\mathsf{T} \mathbf{r}_k}$ \\
% \STATE $\qquad \mathbf{p}_{k+1} := \mathbf{r}_{k+1} + \beta_k \mathbf{p}_k$ \\
% \STATE $\qquad k := k + 1$ \\
% \STATE $\text{end repeat}$ \\
% \STATE $\text{return } \mathbf{x}_{k+1} \text{ as the result}$

% \end{algorithmic}
% \end{algorithm}

\subsubsection{GPU acceleration}

\begin{figure}
\centering
{\includegraphics[width=0.90\hsize]{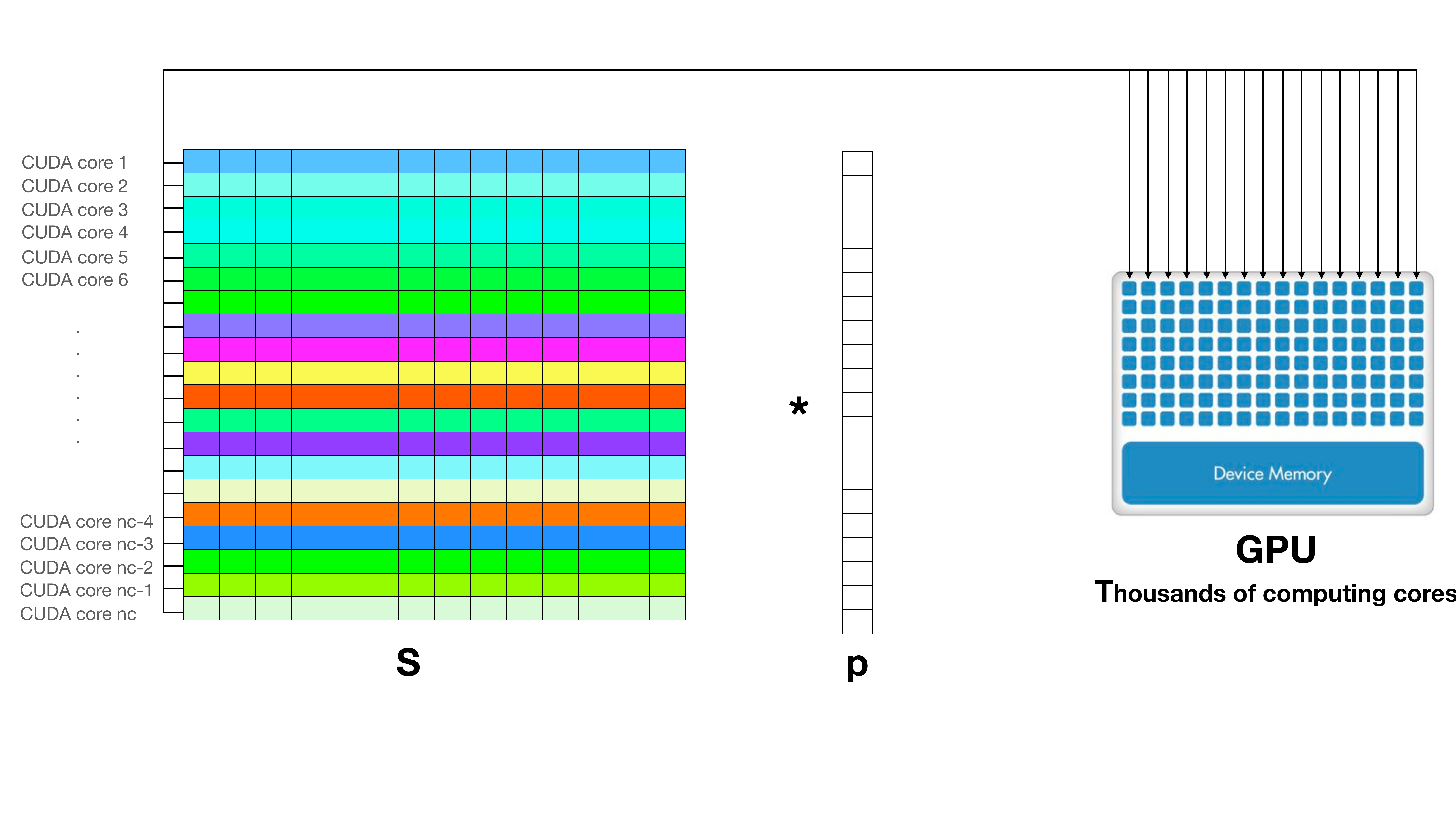}}
\caption{Accelerating matrix-vector multiplication using the GPU involves independent computations for each row of \(\mathbf{S}\) with \(\mathbf{p}\) and each row of \(\mathbf{S}^\top\) with \(\mathbf{s}\). }
\label{fig:GPUcmp}
\end{figure}

The adaptive method described previously decreases the total number of iterations needed by LSQR. In this subsection, we further aim to accelerate each iteration of LSQR by exploiting the high parallel processing power of GPUs. As detailed in Algorithm \ref{algo:lsqr}, the key computational effort in the LSQR algorithm involves the matrix-vector product \(\Smat \mathbf{p}_{k}\) in line 3 and \(\Smat^{\top} \mathbf{s}_{k}\) in line 4 . These operations are well-suited for parallel execution because they primarily consist of basic arithmetic operations like addition, subtraction, multiplication, and division, with minimal complex logic, making them ideal candidates for GPU processing. Utilizing the GPU for matrix-vector multiplication involves parallelizing the computation for each row of \(\Smat\) with \(\mathbf{p}\). This approach distributes the workload across multiple GPU cores, enabling efficient parallel computation and achieving significant speedup. Figure~\ref{fig:GPUcmp} provides an overview of the GPU acceleration framework.

%The adapative approach introduced in the previous section reduces the total number of iteration required to solve a linear system. In this work, we also attemp to speed up the computation of each iteration in the LSQR by leveraging the strong parallel computing capability of GPU. As shown in Algorithm \ref{algo:lsqr},  the primary computational workload in the LSQR algorithm centers around the calculation of the matrix-vector product $\Mmat \mathbf{p}_{k}$ in lines 6 and 8 of Algorithm \ref{algo:lsqr}. These operations are amenable to effective parallelization, primarily involving simple addition, subtraction, multiplication, and division operations without intricate logical processing, making it well-suited for GPU computations. Leveraging the GPU for matrix-vector multiplication entails parallel processing for each row of $\mathbf{M}$ with $\mathbf{p}$. This parallel nature allows the distribution of computations across multiple GPU cores, facilitating efficient parallel processing. Consequently, substantial acceleration can be achieved. Figure~\ref{fig:GPUcmp} presents an overview of the GPU acceleration framework.

To further enhance GPU utilization and performance, we can leverage the independent nature of  $\Smat_{i,j}$. In practice, inverting only one $\Smat_{i,j}$ might underutilize  the GPU's computational power. Since the inversion of   $\Smat_{i,j}$ is independent with the source and frequency indices, we can group multiple $\Smat_{i,j}$ matrices to invert simultaneously, which would be a more efficient strategy. By combining $n_b$ matrices $\Smat_{i,j}$ with identical frequency and achieving comparable convergence rates, the use of iterative solvers for inverting the following block matrix $\mathcal{S}_{k,l} $ becomes advantageous:
%
% Inverting the matrix $\Mmat_{i,j}$ is independent for each source and frequency. Inverting a single $\Mmat_{i,j}$ may not fully utilize the computational capacity of a GPU. To achieve better performance, we can bundle multiple $\Mmat_{i,j}$'s to maximize the usage of GPU resources. Given the roughly similar convergence iterations, it is advantageous to group $n_b \Mmat_{i,j}$‘s with the same frequency. Consequently, we employ iterative solvers to invert the block matrix:
%
%\begin{equation}
%    \mathcal{M}_{k,l} = \begin{bmatrix}
%  \Mmat_{k_1,l} &        0       &  0 & ... & 0 & 0 \\
%        0       & \Mmat_{k_1,l}  &  0 & ... & 0 & 0 \\
%        .       &        .       &  . & ... & . & . \\
%       0       &        .       &  . &     & 0 & \Mmat_{k_{n_b},l}
%  \end{bmatrix}.
%\label{Mmatrix2}
%\end{equation}
%
\begin{equation}
    \mathcal{S}_{k,l} = \begin{bmatrix}
  \Smat_{k_1,l}  \\
  \Smat_{k_1,l} \\
  . \\
  \Smat_{k_{n_b},l}
  \end{bmatrix}.
\label{Mmatrix2}
\end{equation}
In an optimal scenario, employing group inversion techniques can lead to an additional $n_b$ times acceleration in performance. 

\section{Numerical experiments}\label{numerical-examples}

In this section, we evaluate the performance gains and effectiveness of the proposed 3D GPU-accelerated source-free AWRI method, referred to as GPU-SF-AWRI. Since all experiments will utilize source estimation, we will omit the "SF" in GPU-SF-AWRI for simplicity from this point forward. We use the well-known Camembert model and a synthetic example in a Vertical Seismic Profiling (VSP) setting for this assessment.

\subsection{Camembert model}\label{Cambert}
\begin{figure}
\centering
\subfloat[\label{fig:CambertTrue}]{\includegraphics[width=0.50\hsize]{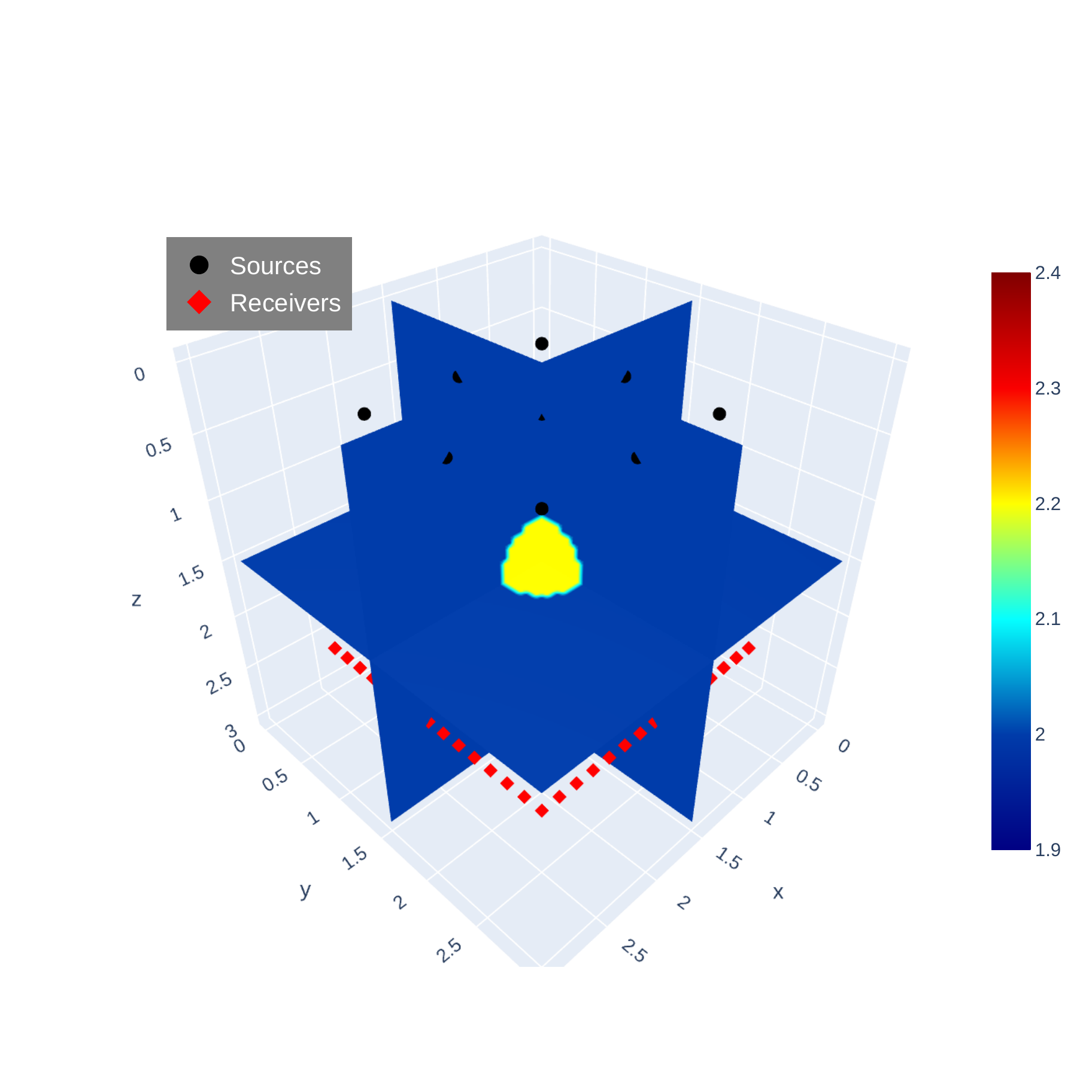}}\
\subfloat[\label{fig:CambertIni}]{\includegraphics[width=0.50\hsize]{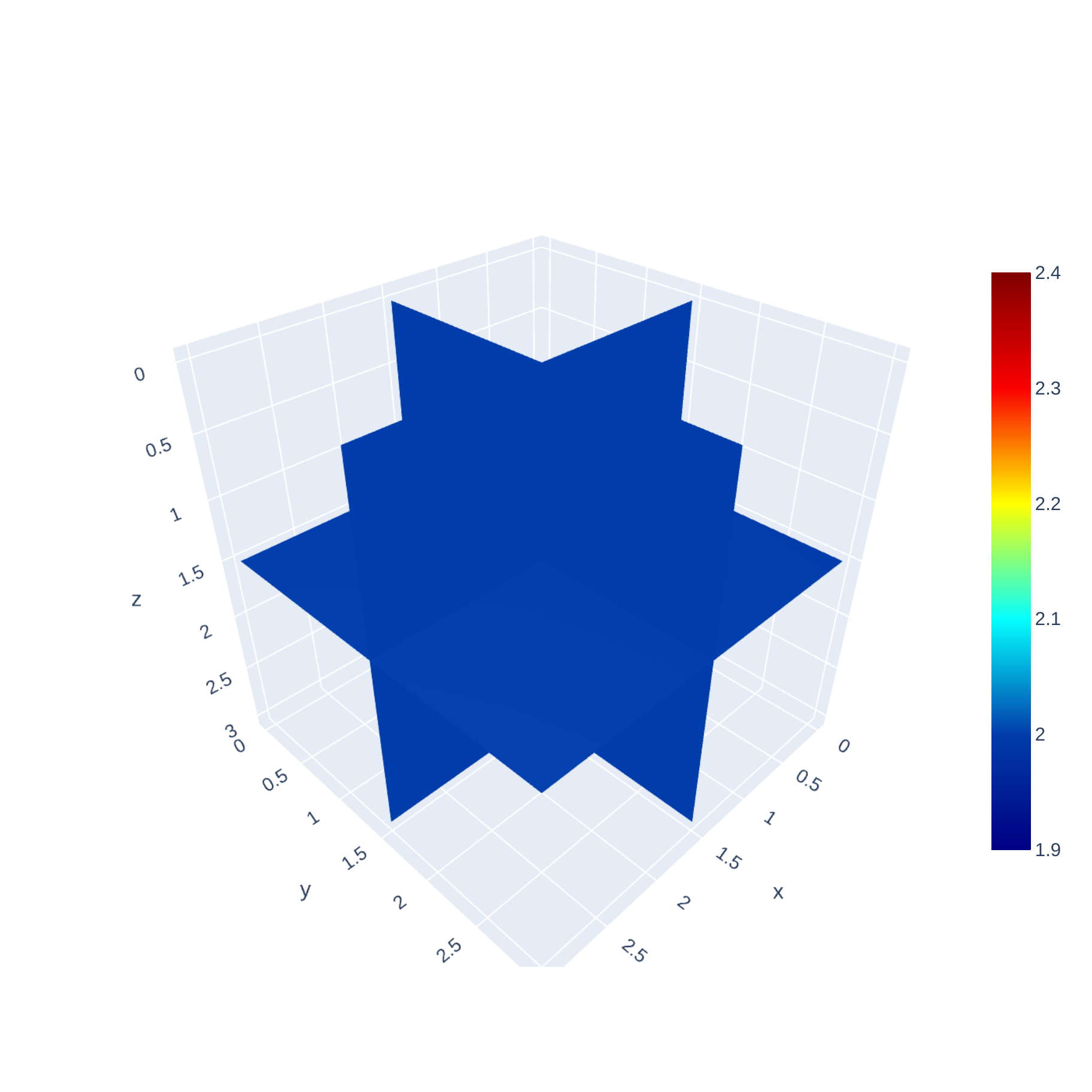}}\
\caption{(a) True model and (b) initial model of the Camembert experiment. }\label{fig:CambertTrueIni}
\end{figure}

We conduct an experiment using the Camembert model to assess the acceleration capabilities of the proposed GPU-AWRI method. As shown in Figure~\ref{fig:CambertTrue}, the model has dimensions of 3 km \(\times\) 3 km \(\times\) 3 km with a grid spacing of \(dx = dy = dz = 0.05\) km. In this setup, we simulate 9 evenly distributed sources located at a depth of \(z = 0.35\) km, and collect data with 2025 receivers uniformly placed at a depth of \(z = 2.3\) km. The source function used is a Ricker wavelet with a central frequency of \(f = 10\) Hz.

\subsubsection{Evaluation of GPU Acceleration Efficiency}

We first conduct a comprehensive analysis of the speed performance of GPU-AWRI by comparing the computation times required for a single misfit function calculation using CPU-based 3DWRI, GPU-based 3DWRI without the grouping strategy, and GPU-based 3DWRI with the grouping strategy. We conduct the comparison using data at the frequency of 5 Hz and the model shown in Figure \ref{fig:CambertIni}. The penalty parameter is selected as $\lambda=1\times10^{-1}$. The tolerance for LSQR is fixed with $\epsilon = 1\times 10^{-6}$. The assessment is carried out on a machine equipped with an NF5468M5 CPU, featuring 2 Intel Xeon4214R (2.4GHz/12-Core) processors, and an NVIDIA A100 Tesla GPU. In both approaches, the iterative solver is halted upon reaching a relative residual of $1\times10^{-6}$. For 3D GPU-WRI with the grouping strategy, we bundle 4 sources together to maximize GPU computational capacity utilization. The results, presented in Table~\ref{tab:CMPTime}, are noteworthy. Compared to CPU-based 3D WRI, 3D GPU-WRI without the grouping strategy shows an almost 80-fold speedup (7424 seconds vs 93 seconds). Furthermore, with the proposed grouping strategy, 3D GPU-WRI demonstrates an impressive 195$\times$ speedup (7424 seconds vs 38 seconds) compared to its CPU-based counterpart. This substantial improvement in performance highlights the GPU-accelerated method's ability to calculate a single misfit function a hundred times faster, offering significant advantages for 3D WRI applications.

\begin{table}
    \centering
    \begin{tabular}{c|ccc}
        \hline
                       &3D CPU-WRI & 3D GPU-WRI   & 3D GPU-WRI \\
                       &  & without grouping & with grouping \\
         \hline
         Time (s) &7424 seconds & 93 seconds & 38 seconds\\
         \hline
    \end{tabular}
    \caption{Comparison of computational time between WRI using CPU, GPU, and GPU with grouping strategy.}
    \label{tab:CMPTime}
\end{table}

\begin{figure}
\centering
{\includegraphics[width=0.800\hsize]{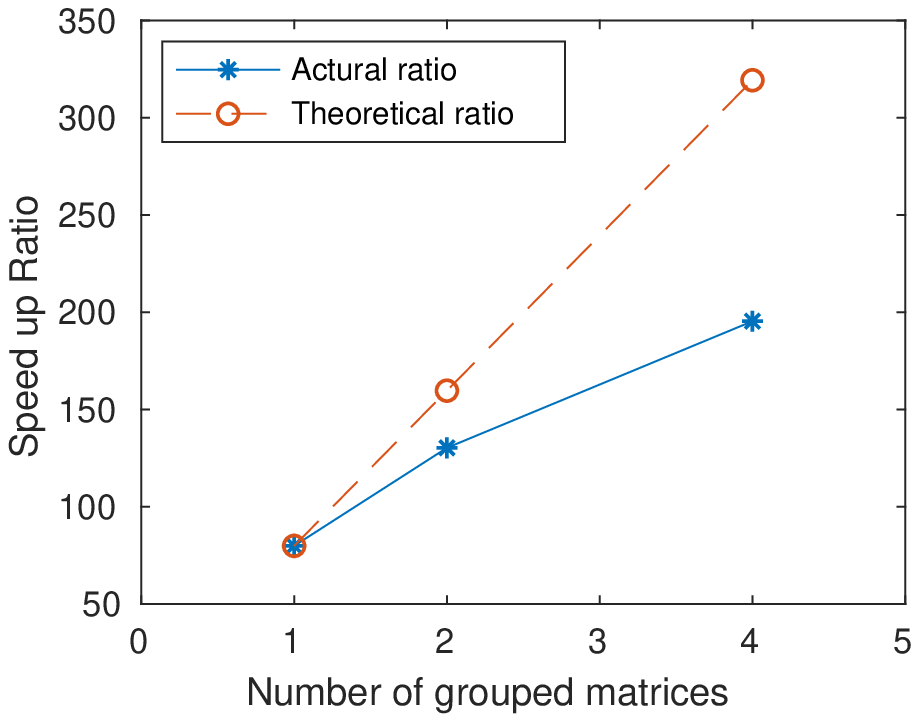}}
% \subfloat[\label{fig:MDEcmp}]{\includegraphics[width=0.400\hsize]{./Figure4a.eps}}
\caption{Acceleration ratios corresponding to the choices of $n_b$ = 1, 2, and 4.}\label{fig:GPUgrouping}
\end{figure}

Furthermore, we assess the acceleration performance of the proposed grouping strategy by evaluating a misfit function for various numbers of grouped matrices, denoted as \( n_b \) = 1, 2, and 4. Figure~\ref{fig:GPUgrouping} presents a comparison between the observed acceleration ratio and the expected theoretical ratio. As anticipated, the actual speed-up ratio falls short of the theoretical ratio. Notably, as \( n_b \) increases, indicating a higher GPU load, the speed-up performance diminishes. This is evident from the comparison of speed-up ratios for \( n_b = 2 \) and \( n_b = 4 \).

\subsubsection{Evaluation of GPU-AWRI}

We further evaluate the performance of the proposed 3D GPU-AWRI method by conducting four inversions under different configurations: (1) WRI with a constant small tolerance of \(\epsilon = 1 \times 10^{-6}\); (2) WRI with a constant large tolerance of \(\epsilon = 4 \times 10^{-4}\); (3) AWRI with an adaptive tolerance starting from \(\epsilon_0 = 4 \times 10^{-4}\); and (4) AWRI with  TV regularization. All four experiments leverage GPU acceleration and use 50 l-BFGS iterations. We fine-tuned the penalty parameter to $\lambda=1\times10^{-1}$ and the TV weighting parameter to $\gamma=1 \times 10^{-3}$. All inversions are performed using an incorrect initial source function centered at 10 Hz with a time delay of 0.11 seconds.

\begin{figure}
\centering
\subfloat[\label{fig:Cam1e6}]{\includegraphics[width=0.500\hsize]{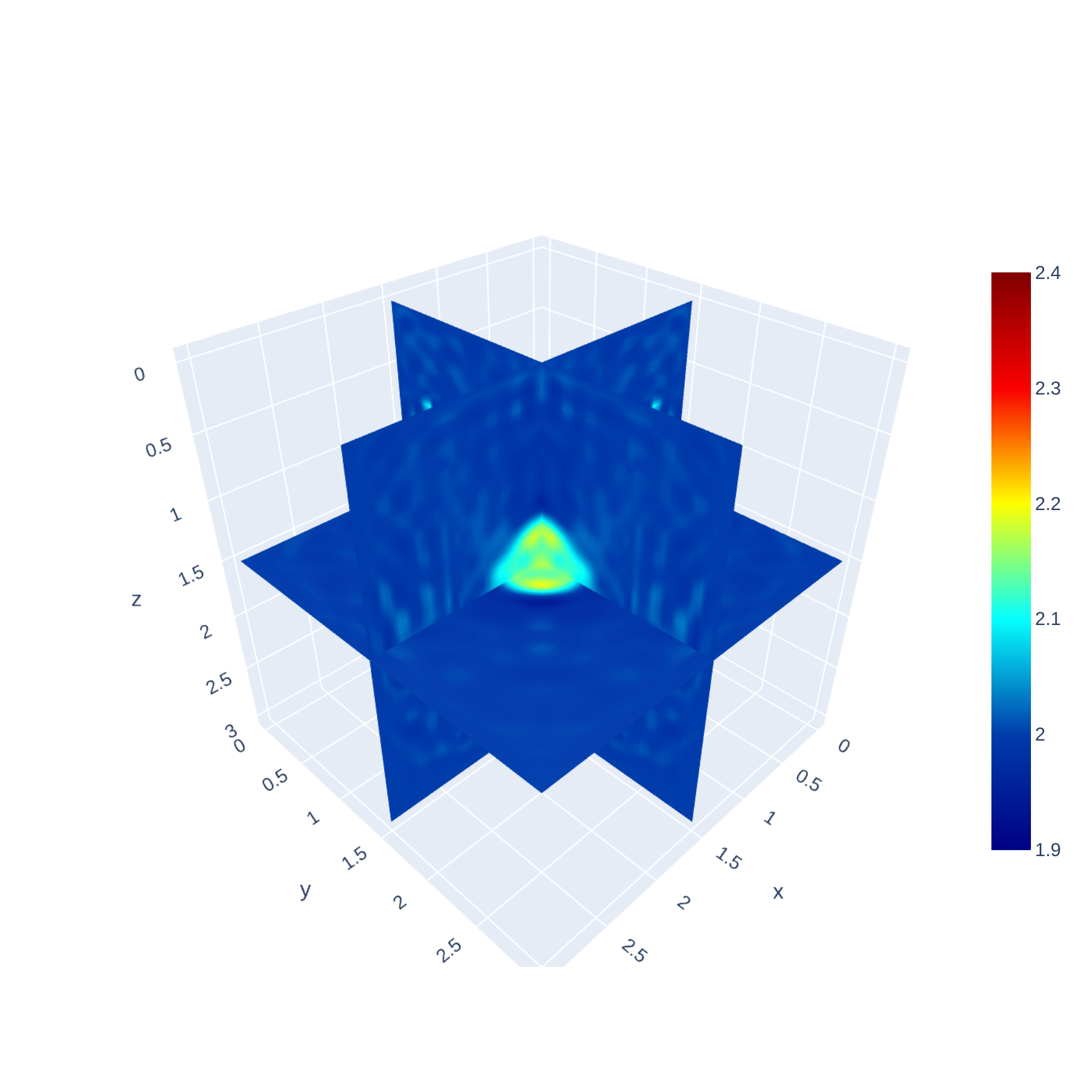}}
\subfloat[\label{fig:Cam4e4}]{\includegraphics[width=0.500\hsize]{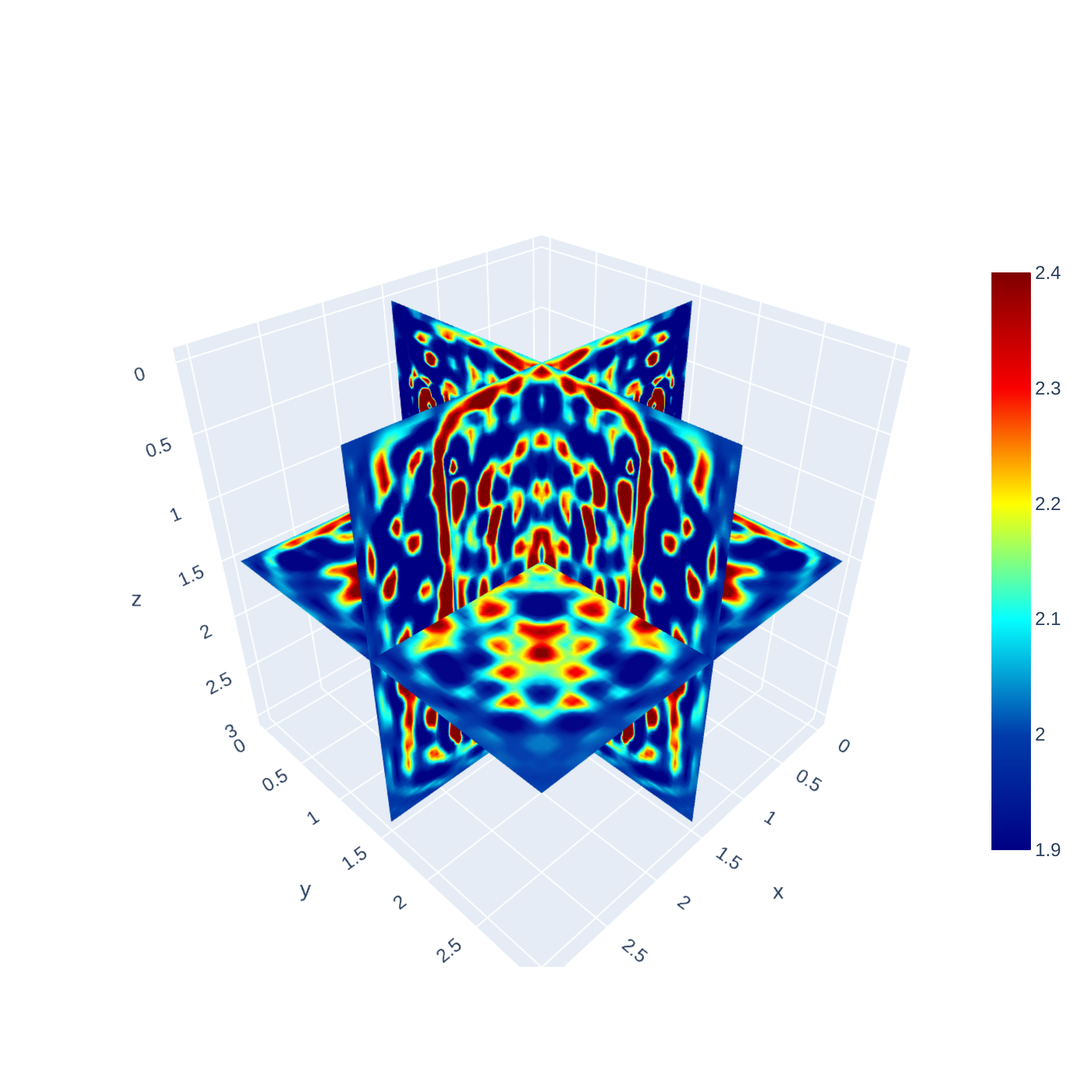}}\\
\subfloat[\label{fig:CamAWRI}]{\includegraphics[width=0.500\hsize]{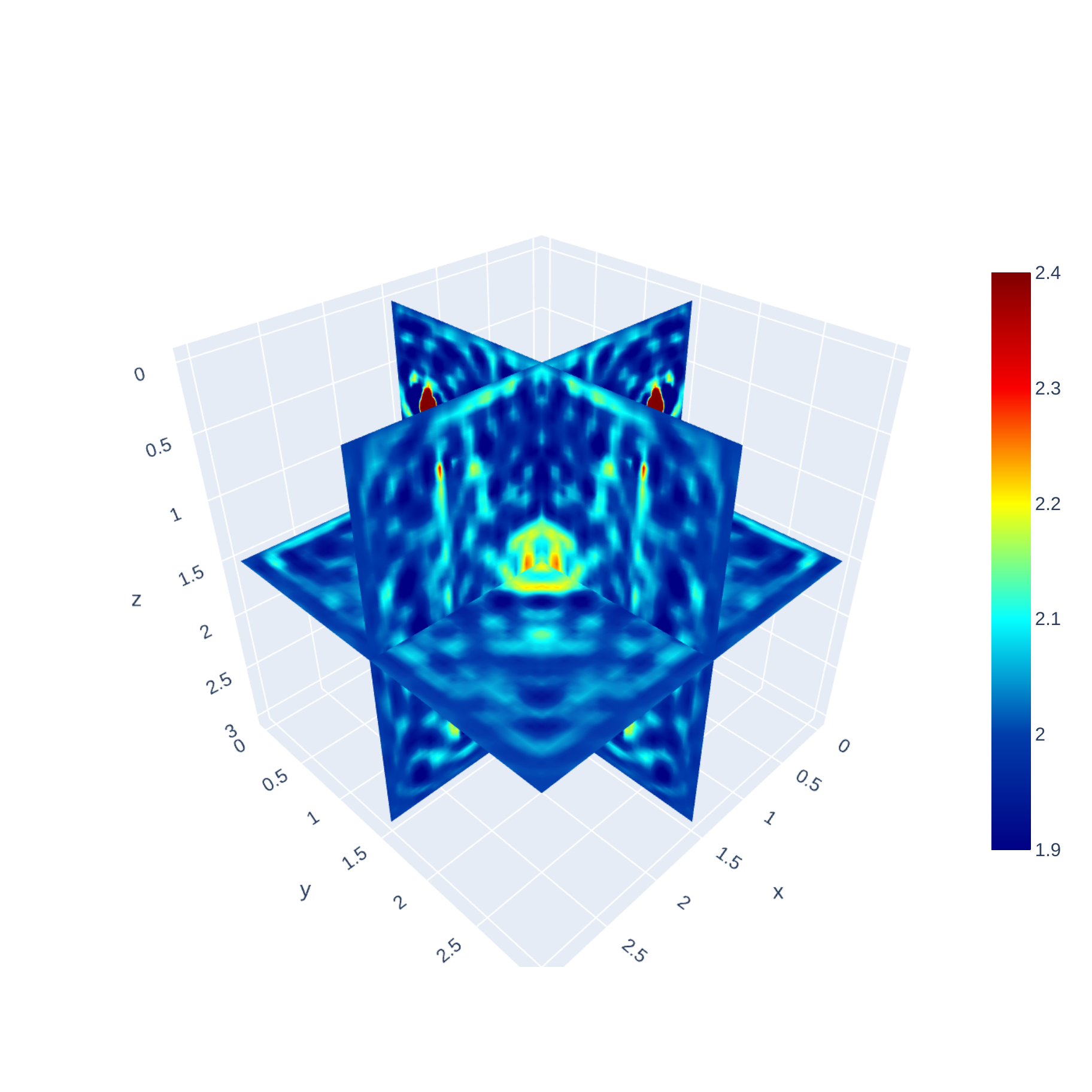}}
\subfloat[\label{fig:CamAWRITV}]{\includegraphics[width=0.500\hsize]{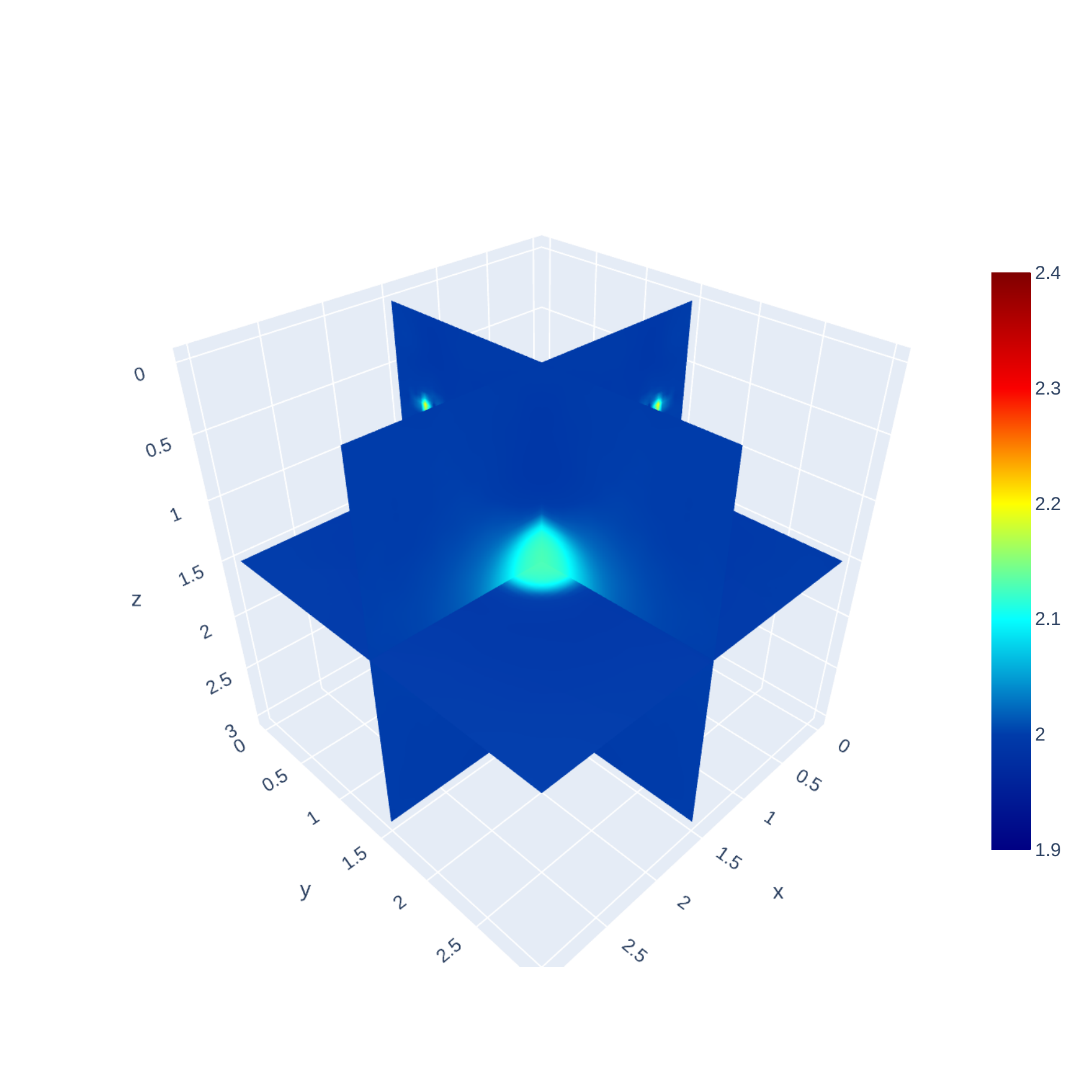}}
% \subfloat[\label{fig:MDEcmp}]{\includegraphics[width=0.400\hsize]{./Figure4a.eps}}
\caption{Results of (a) WRI with tolerance  \(\epsilon = 1 \times 10^{-6}\), (b) WRI with tolerance of \(\epsilon = 4 \times 10^{-4}\), (c) AWRI, and (d) AWRI with TV regularization.}\label{fig:AWRIresult}
\end{figure}

Figure \ref{fig:AWRIresult} presents the outcomes of the four different methods. Table \ref{tab:MECamInv} show the relative model error $\frac{\|\mvec_{f}-\mvec_{t}\|}{\|\mvec_{t}\|}$ between the true model $\mvec_{t}$ and the final results $\mvec_{f}$, and the corresponding computational times for each method, respectively. It is evident that the WRI method with $\epsilon = 1 \times 10^{-6}$ yields the most accurate inversion, with the smallest relative model error of $0.56\%$, but it also uses the highest computational time at 0.56 hours. Conversely, WRI with $\epsilon = 4 \times 10^{-4}$ achieves the fastest computation but results in the largest relative model error of $12.53\%$. Both AWRI and AWRI with TV significantly reduce computation time to nearly half of that required by WRI with $\epsilon = 1 \times 10^{-6}$, while producing much better results than WRI with $\epsilon = 4 \times 10^{-4}$. Their relative model errors are $3.51\%$ and $0.62\%$, respectively. As expected, employing adaptive tolerance impacts the inversion accuracy, but incorporating TV regularization helps maintain accuracy. The final relative model error of AWRI with TV reaches the level of WRI with $\epsilon = 1 \times 10^{-6}$.
 
\begin{table}
    \centering
    \begin{tabular}{c|cccc}
        \hline
                                         & WRI                                          & WRI                                          & AWRI & AWRI  \\
                                         & ($\epsilon = 1 \times 10^{-6}$) & ($\epsilon = 4 \times 10^{-4}$) &           & w/ TV  \\
        \hline
        Model Error($\%$)   & 0.56                                          & 12.53                                        & 3.51  & 0.62\\
        Time (h)                   & 0.56                                          & 0.09                                         & 0.33  & 0.30 \\     
         \hline
    \end{tabular}
    \caption{Comparison of relative model errors ($\%$) and computational time (hour) for the four methods.}
    \label{tab:MECamInv}
\end{table}

%\begin{table}
%    \centering
%    \begin{tabular}{|c|c|c|c|c|}
%        \hline
%                             & WRI                                          & WRI                                          & AWRI & AWRI  \\
%                             & ($\epsilon = 1 \times 10^{-6}$) & ($\epsilon = 4 \times 10^{-4}$) &           & w/ TV  \\
%        \hline
%        Time             & 0.56                                          & 0.09                                         & 0.33  & 0.30 \\
%        (hour)           &                                                  &                                                 &           &  \\
%         \hline
%    \end{tabular}
%    \caption{Computational time comparison of four inversion methods.}
%    \label{tab:CMPTimeCamInv}
%\end{table}

\begin{figure}
\centering
{\includegraphics[width=0.800\hsize]{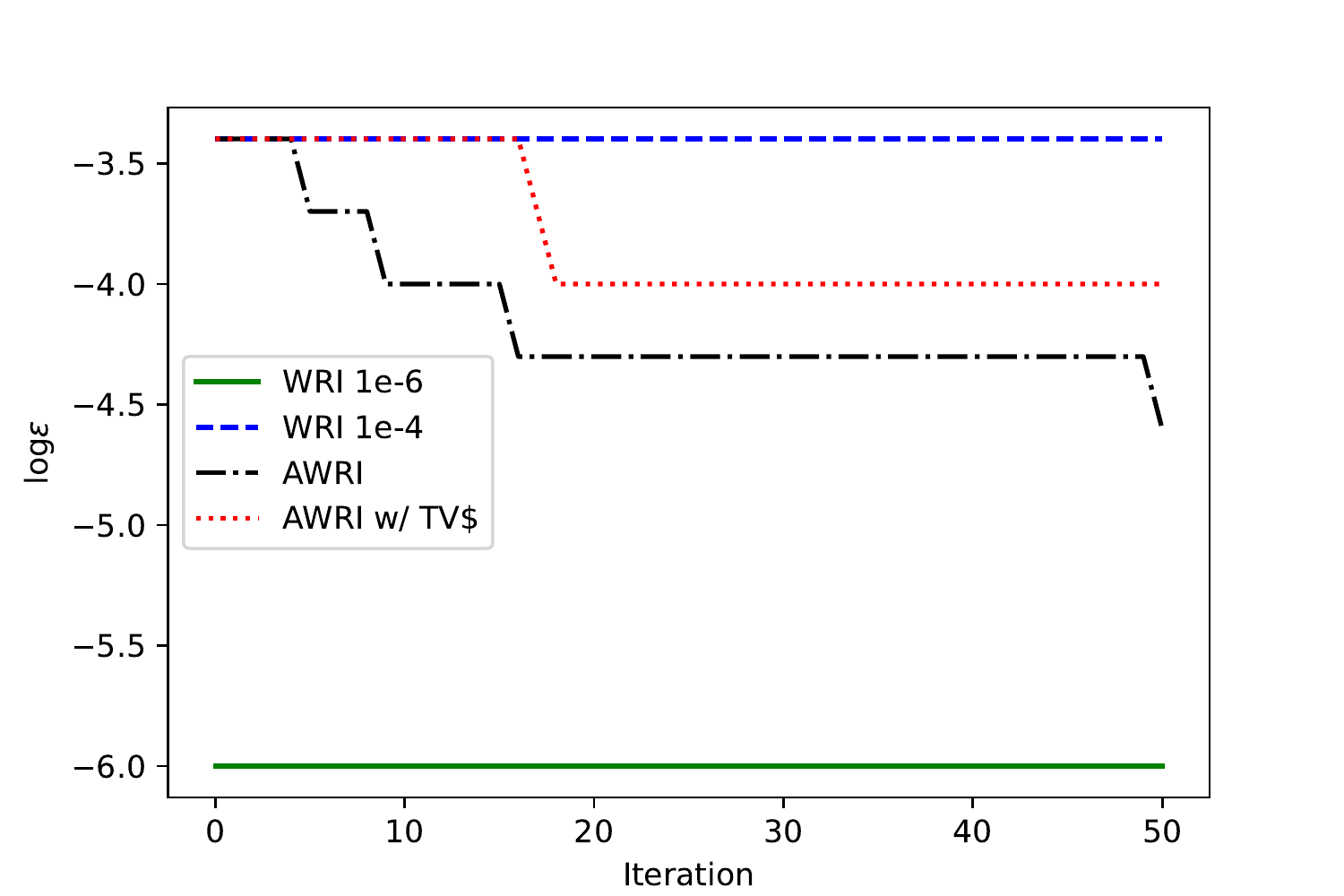}}
% \subfloat[\label{fig:MDEcmp}]{\includegraphics[width=0.400\hsize]{./Figure4a.eps}}
\caption{Tolerance v.s. Iterations. }\label{fig:CamTolCmp}
\end{figure}

Figure \ref{fig:CamTolCmp} depicts the tolerance progression for the four methods. AWRI and AWRI with TV regularization exhibit a gradual reduction in tolerance, effectively decreasing computational time. Additionally, TV regularization mitigates artifacts introduced by less accurate wavefields, enabling AWRI to operate at higher tolerance levels. This optimization further reduces computational time, as shown in Table \ref{tab:MECamInv}.

Finally, in Table \ref{tab:SrcCam}, we compare the true source \(\alpha_{t}\), the initial source \(\alpha_{0}\), and the final source \(\alpha_{f}\) estimated by AWRI. Initially, the source function exhibits a significant relative error of 34.4\% \(\left(\frac{\|\alpha_{0} - \alpha_{t}\|}{\|\alpha_{t}\|}\right)\). As AWRI progresses, the proposed source estimation method gradually refines the source function, ultimately achieving a relatively accurate source with a reduced relative error of 2.7\%.

\begin{table}
    \centering
    \begin{tabular}{c|ccc}
        \hline
                                      &   5Hz                                          & 6Hz                 & Relative error ($\%$) \\
        \hline
        $\alpha_{t}$              &  -10 + 0i				        &-8.1 + 5.9i        &  -\\
        $\alpha_{0}$            &  -9.51 + 3.1i                              & -5.36 + 8.44i   &   34.4\\       
        $\alpha_{f}$              & -10.15-0.28i                             &  -8.28+5.74i    &  2.7 \\     
         \hline
    \end{tabular}
    \caption{Comparison of the true source \(\alpha_{t}\), the initial source \(\alpha_{0}\), and the final source \(\alpha_{f}\) estimated by AWRI.}
    \label{tab:SrcCam}
\end{table}

\subsection{VSP numerical model}\label{VSPNM}

\begin{figure}
\centering
\subfloat[\label{fig:VSPNMTrue}]{\includegraphics[width=0.500\hsize]{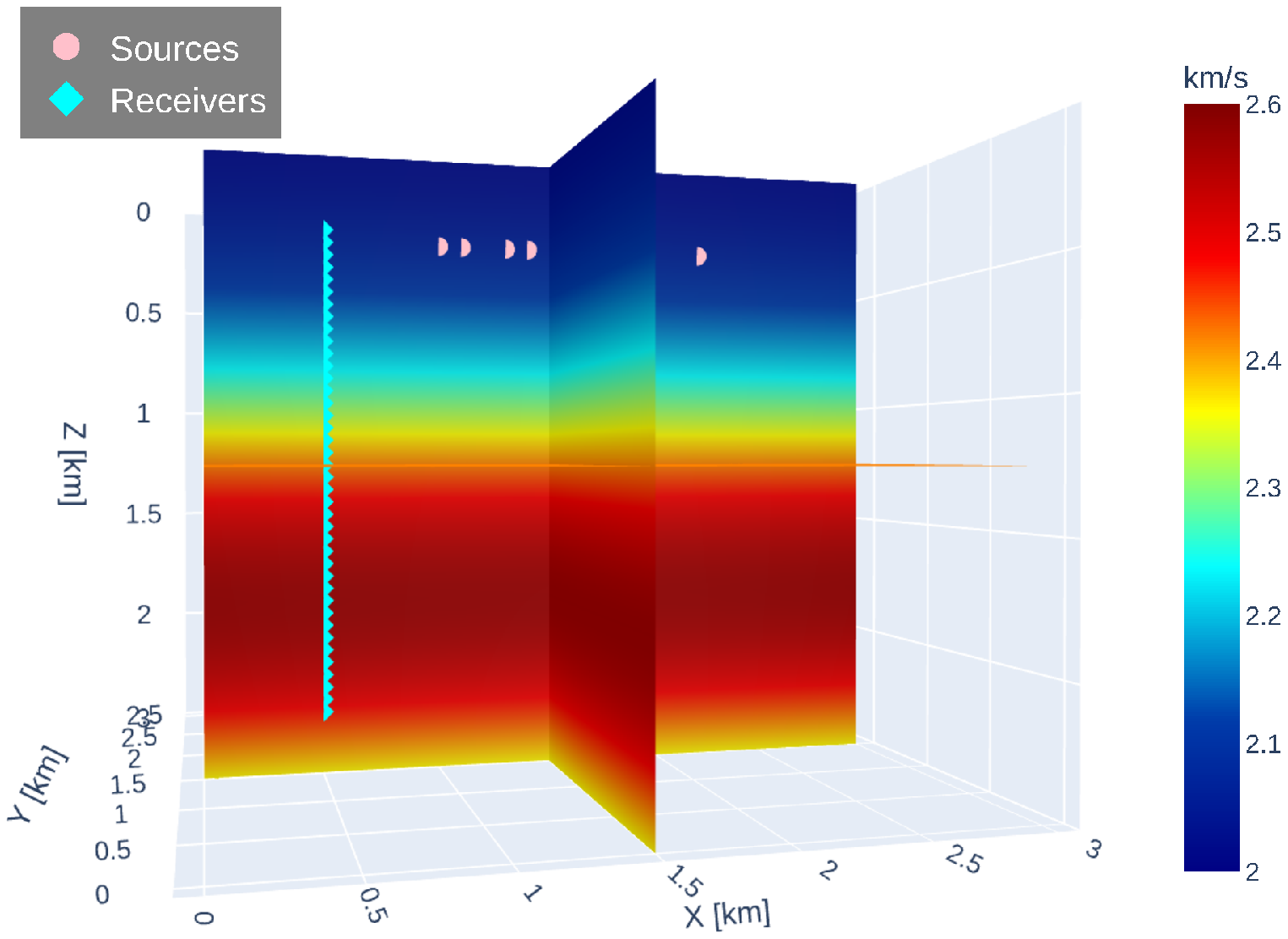}}\\
\subfloat[\label{fig:VSPNMIni}]{\includegraphics[width=0.500\hsize]{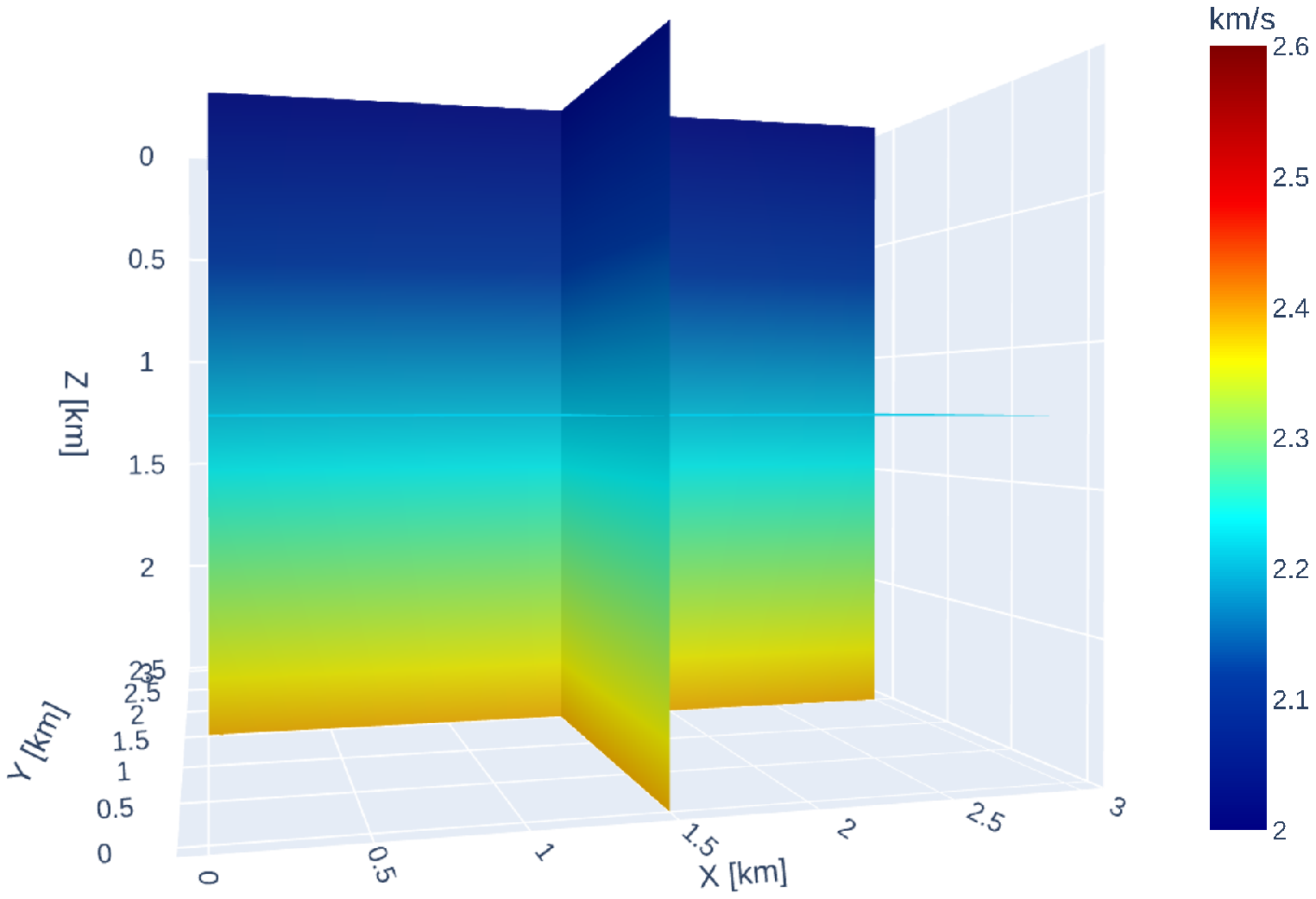}}\\
% \subfloat[\label{fig:MDEcmp}]{\includegraphics[width=0.400\hsize]{./Figure4a.eps}}
\caption{(a) True model and (b) initial model of the VSP numerical example.}\label{fig:VSPNMModels}
\end{figure}

We first performed a 3D VSP numerical experiment to assess the efficacy of GPU-AWRI in mitigating local minima. The model dimensions are 3 km \(\times\) 3 km \(\times\) 2.5 km, as depicted in Figure \ref{fig:VSPNMTrue}, with variation occurring only along the depth axis. We positioned 8 equally spaced sources from \(x = 1\) km to \(x = 2.2\) km, aligned along \(y = 1.5\) km at a depth of \(z = 0.35\) km. Additionally, 50 receivers were placed at intervals of 0.05 km in a well position at \([x, y] = [0.5\) km, \(1.5\) km]. Data was simulated using a Ricker wavelet centered at \(f = 10\) Hz. 

In this paper, we compare three types of inversions: conventional FWI, WRI with a constant tolerance \(\epsilon = 1 \times 10^{-6}\), and GPU-AWRI with TV regularization. For GPU-AWRI, we start with the tolerance $\epsilon=1 \times 10^{-4}$. The initial model used in the inversions is shown in Figure \ref{fig:VSPNMIni}, and data was processed at a frequency of 5 Hz. Although the true model varies only along the depth axis, we performed 3D inversions to infer the 1D velocity profile. For WRI and AWRI, we adjusted the penalty parameter \(\lambda = 1 \times 10^{-2}\). For GPU-AWRI with TV regularization (AWRI-TV), we fine-tuned the TV weighting parameter \(\gamma = 1 \times 10^{-6}\).

\begin{figure}
\centering
\subfloat[\label{fig:VSPNMFWI}]{\includegraphics[width=0.500\hsize]{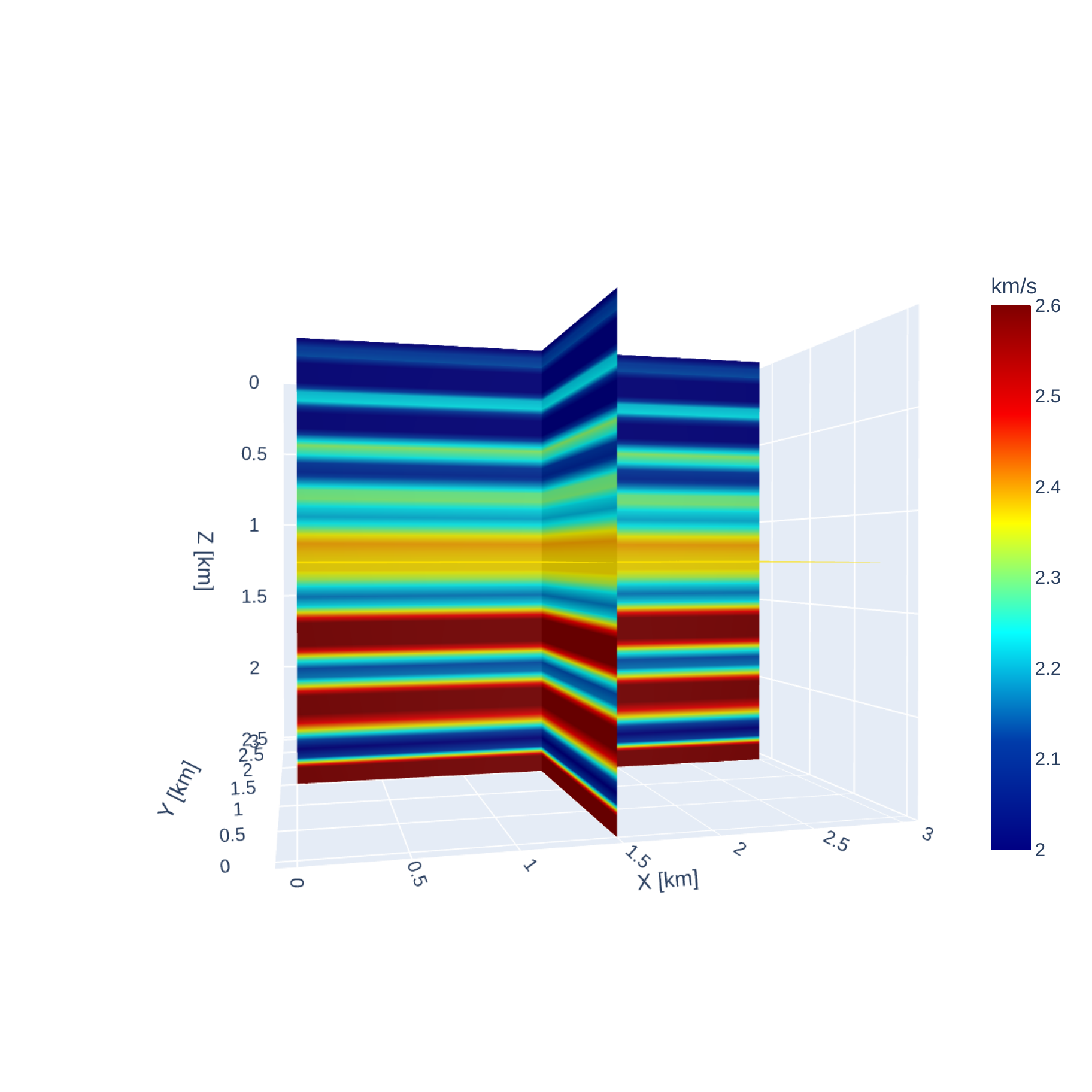}}\\
\subfloat[\label{fig:VSPNMWRI}]{\includegraphics[width=0.500\hsize]{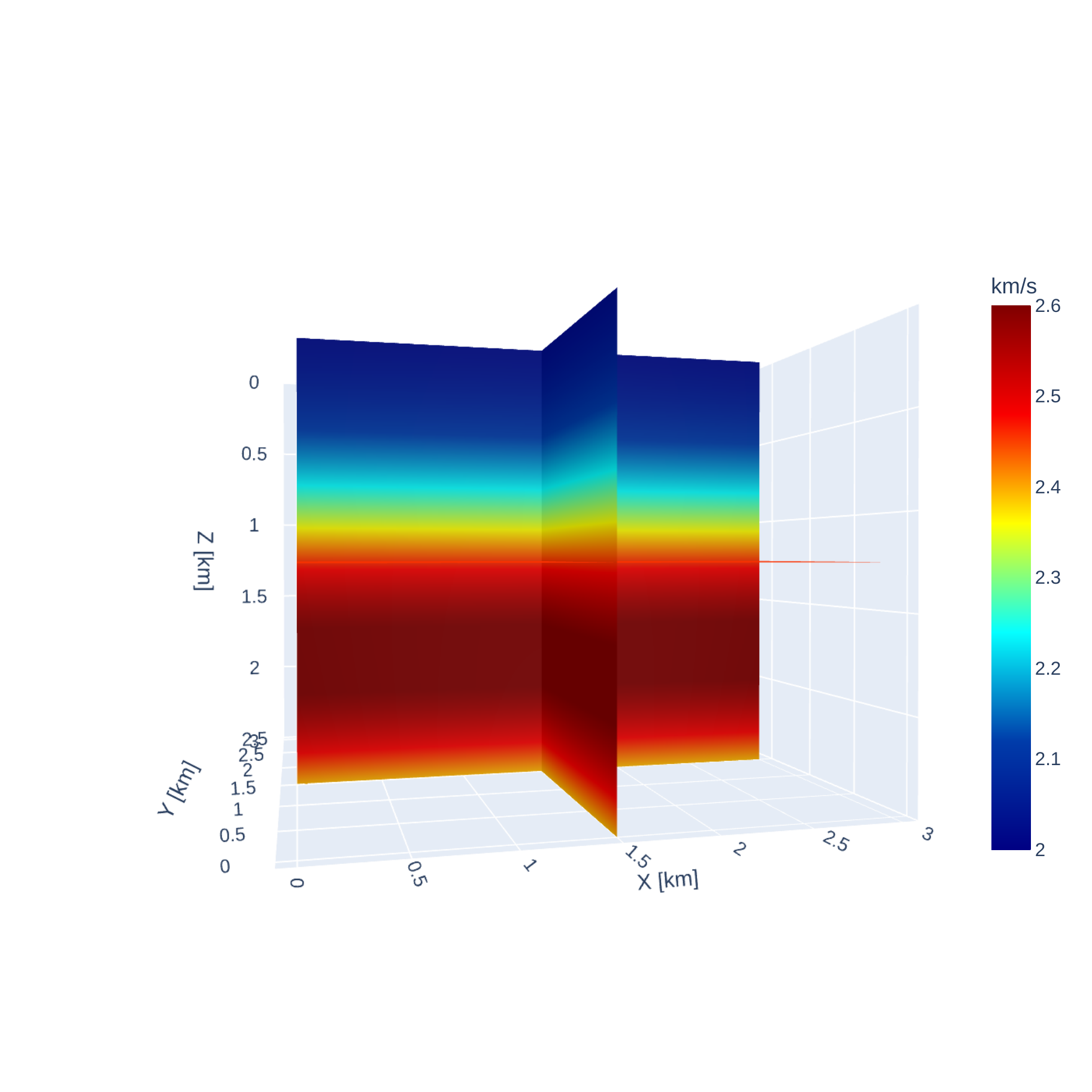}}
\subfloat[\label{fig:VSPNMWRI}]{\includegraphics[width=0.500\hsize]{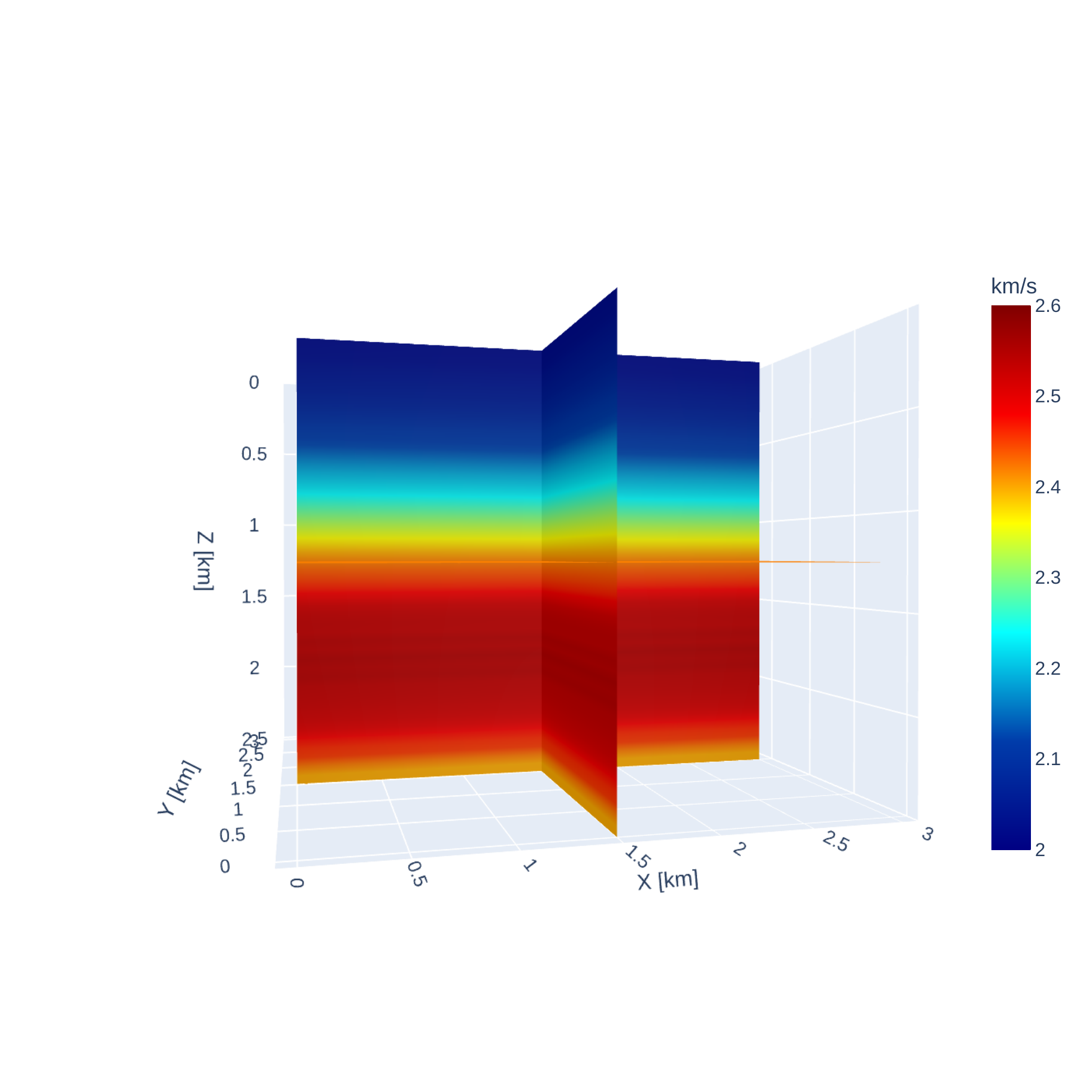}}
% \subfloat[\label{fig:MDEcmp}]{\includegraphics[width=0.400\hsize]{./Figure4a.eps}}
\caption{Inversion results of (a) FWI, (b) WRI and (c)  AWRI-TV of the VSP numerical example.}\label{fig:VSPNMResults}
\end{figure}

\begin{figure}
\centering
{\includegraphics[width=0.800\hsize]{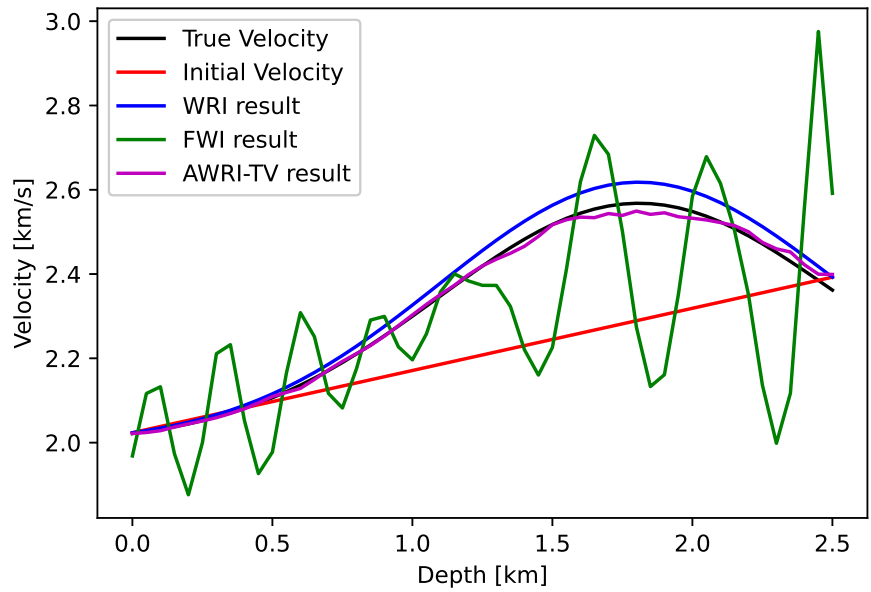}}
% \subfloat[\label{fig:MDEcmp}]{\includegraphics[width=0.400\hsize]{./Figure4a.eps}}
\caption{Comparison of 1D Velocity Profiles Between the True and initial Models and the Inverted Results of FWI, WRI and AWRI-TV.}\label{fig:VSPNM1DVel}.
\end{figure}

Figure~\ref{fig:VSPNMResults} depicts the inverted models obtained by FWI, WRI, and AWRI-TV, while Figure~\ref{fig:VSPNM1DVel} compares the 1D velocity profiles among the true model, initial model, and results from FWI, WRI, and AWRI-TV. Table~\ref{tab:MEVSPNInv} summarizes the final relative model errors of the three methods. Clearly, FWI falls into local minima with a relative model error of $8.3\%$. In contrast, both WRI and AWRI-TV effectively recover the velocity model, achieving relative model errors of $1.4\%$ and $0.5\%$, respectively. Notably, AWRI-TV inherits WRI's ability to mitigate local minima, and the addition of TV helps maintain inversion accuracy.

\begin{table}
    \centering
    \begin{tabular}{c|ccc}
        \hline
                                          & FWI                                          & WRI                                          & AWRI w/ TV  \\
        \hline
        Model Error($\%$)     & 8.3                                          & 1.4                                             & 0.5\\
         \hline
    \end{tabular}
    \caption{Relative model error $\frac{\|\mvec_{f}-\mvec_{t}\|}{\|\mvec_{t}\|}$ comparison of  three inversion methods.}
    \label{tab:MEVSPNInv}
\end{table}

Table \ref{tab:SrcVSPNM} compares the true source \(\alpha_{t}\), initial source \(\alpha_{0}\), and final source \(\alpha_{f}\) estimated by AWRI. As anticipated, AWRI progressively refines the source function, reducing the relative error from 31.4\% to 2.7\%.

\begin{table}
    \centering
    \begin{tabular}{c|cc}
        \hline
                                      &   5Hz                             &  Relative Error ($\%$)              \\  
        \hline
        $\alpha_{t}$              &  -10 + 0i			  &   -	        \\
        $\alpha_{0}$            &  -9.51 + 3.1i               &   31.4           \\       
        $\alpha_{f}$              & -9.83 - 0.25i              &  2.9             \\     
         \hline
    \end{tabular}
    \caption{Comparison of the true source \(\alpha_{t}\), the initial source \(\alpha_{0}\), and the final source \(\alpha_{f}\) estimated by AWRI.}
    \label{tab:SrcVSPNM}
\end{table}

\begin{figure}
\centering
{\includegraphics[width=0.800\hsize]{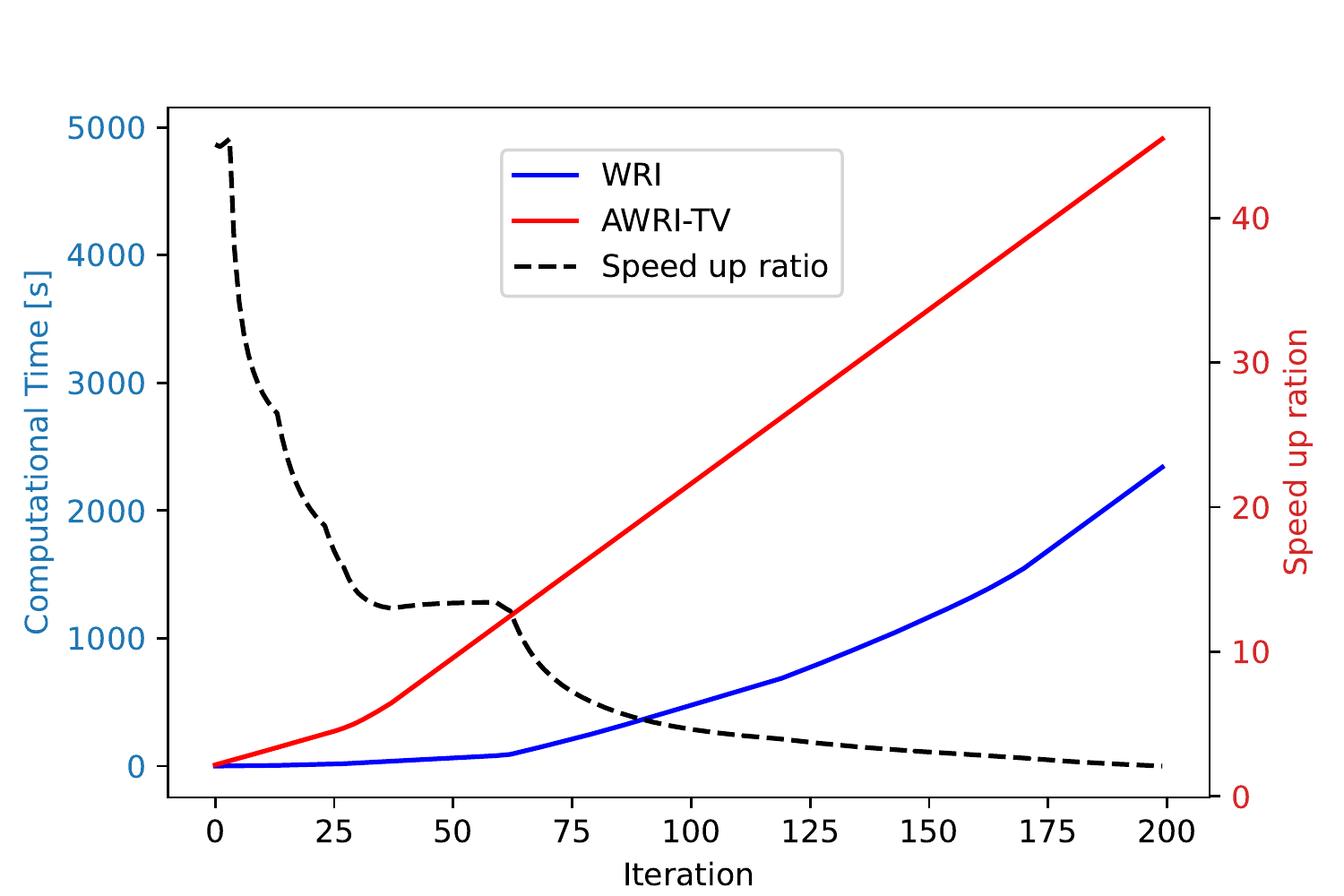}}
% \subfloat[\label{fig:MDEcmp}]{\includegraphics[width=0.400\hsize]{./Figure4a.eps}}
\caption{Computational time and Speed up ratio history. }\label{fig:VSPNCmpCmp}
\end{figure}

Figure \ref{fig:VSPNCmpCmp} presents the computational time history for both WRI and AWRI-TV, along with the corresponding speed-up. In the initial stages of the inversion, AWRI-TV achieves nearly a 50-fold speed-up by utilizing a relatively high tolerance. As the inversion progresses and stricter tolerances are required, the speed-up ratio decreases. Ultimately, AWRI-TV achieves a 2-fold speed-up compared to WRI, while delivering results with lower relative model error.

\section{Physical modeling experiment}\label{PM}
Following the numerical example, we gauge the practicality of the proposed 3D GPU-AWRI for real-world applications by employing it on an experimental physical modeling dataset under a VSP survey configuration. The controlled physical modeling experiment is an ideal method for validating the numerical approach by providing a realistic dataset from a known ground truth in comparison to field surveys. Compared with the numerical modeling experiment, the physical modeling experiment leverages the advantage that the recorded seismic data has no direct connection to the inverse solver, ensuring unbiased validation. To create an analog of the geological formation, we first construct a downscaled physical model using materials with elastic properties similar to those of the actual geological formations. Using ultrasonic transducers with a center frequency of 300 kHz, we adopt a scale of 1:10,000 for frequency downscaling, which results in time and space upscaling in the physical modeling experiment. Consequently, the central frequency of 30 kHz is equivalent to 30 Hz after scaling, and a 1-mm model dimension corresponds to a 10-m field dimension. In this study, we utilized plexiglass, glass, and rock salt as analogs for geological formations with varying elastic properties. Table~\ref{tab:pm_par} lists the elastic properties of the employed materials and the respective depths after up-scaling. We design a walkaway-VSP survey configuration that employs sources excited at the model surface. The receiving transducer is positioned against the model's side and is maneuvered along the vertical edge by moving gantries, serving as an analog for borehole receivers.

\begin{table*}[!t]
\caption{Elastic parameters of materials employed in the physical modeling.}
  \centering
	\begin{tabular}{c|c|cccc}
	\hline
	{ \textbf{Layer}} & { \textbf{Physical model material}} & { \textbf{$V_\mathrm{P}$ (m/s)}} & { \textbf{$V_\mathrm{S}$ (m/s)}} & { \textbf{Density (kg/m$^3$)}} & { \textbf{Depth (m)}} \\ \hline
	{1}                 & {Plexiglass}        & {2740}              & {1380}              & {1180}                & {0-846.8}                           \\
	{2}                 & { Glass}             & {5830}              & {3340}              & { 2700}                 & {846.8-1100.5}                      \\
	{3}                 & {Rock salt}         & {4600}              & {2630}              & { 2160}                & {1100.5-1522.5}                     \\
	{N/A}                & {Supersaturated brine}             & {1846}              & {N/A}                & {1202}               & {N/A}                                \\ \hline                                            
	\end{tabular}
    \label{tab:pm_par}
\end{table*}

\subsection{Experimental setup}
\begin{figure}[!htbp]
\centering
\subfloat[\label{fig:pm_side}]{\includegraphics[width=0.500\hsize]{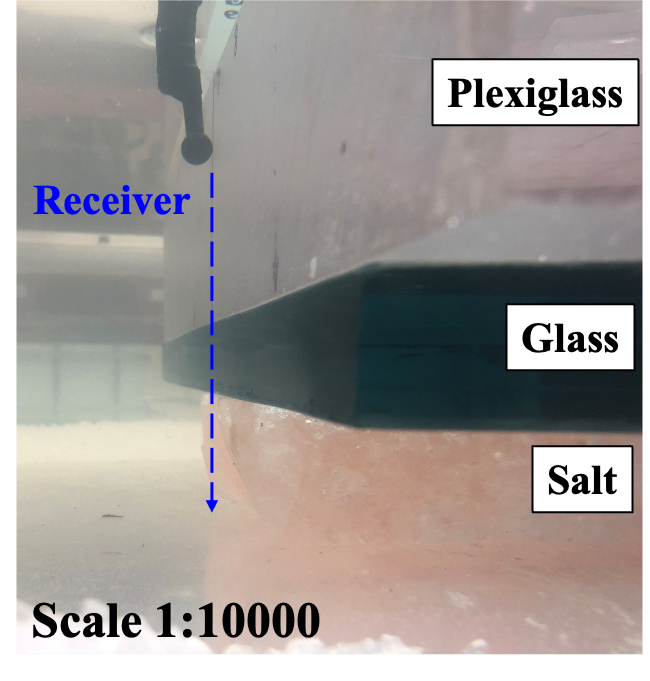}}\\
\subfloat[\label{fig:pm_top}]{\includegraphics[width=0.500\hsize]{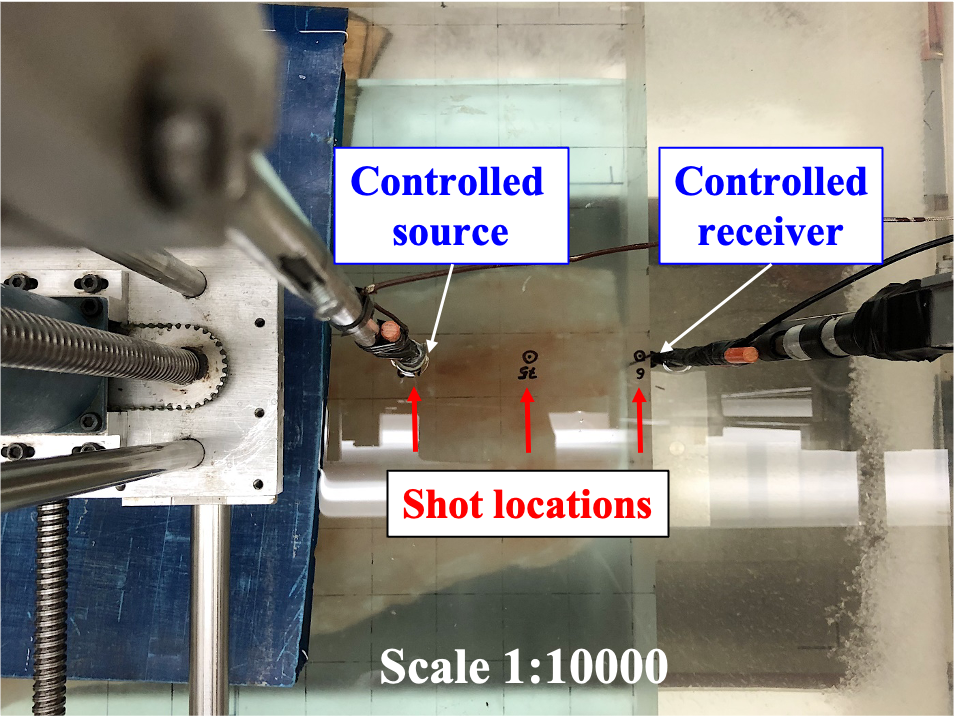}}
\caption{Experimental setup for physical modeling experiment from the (a) side and (b) top view.}
\label{pm_setup}
\end{figure}

The physical modeling experiments were undertaken at the Allied Geophysical Laboratories at the University of Houston. Figure~\ref{fig:pm_side} and Figure~\ref{fig:pm_top} display the experimental setup from the side and top view, respectively. To create a walkaway VSP survey geometry, we designed three aligned shot points located from near to far offset at the top of the model. The source and receiver transducers are attached to moving gantries controlled by a computer program configured with the designed acquisition geometry. To ensure proper coupling between the ultrasonic transducers and the model, the model was placed in a tank filled with supersaturated brine, which does not affect the physical properties or shape of the model materials.
For the VSP data acquisition, the receiver transducer was moved along the vertical side of the model at 10-m depth intervals, covering a total depth range of 1400 m, after scaling. This was repeated for all three shot locations, which correspond to scaled depths of 60, 750, and 1500 m.

\subsection{Wavefield preparation}
\begin{figure}[!htbp]
\centering
\includegraphics[width=\hsize]{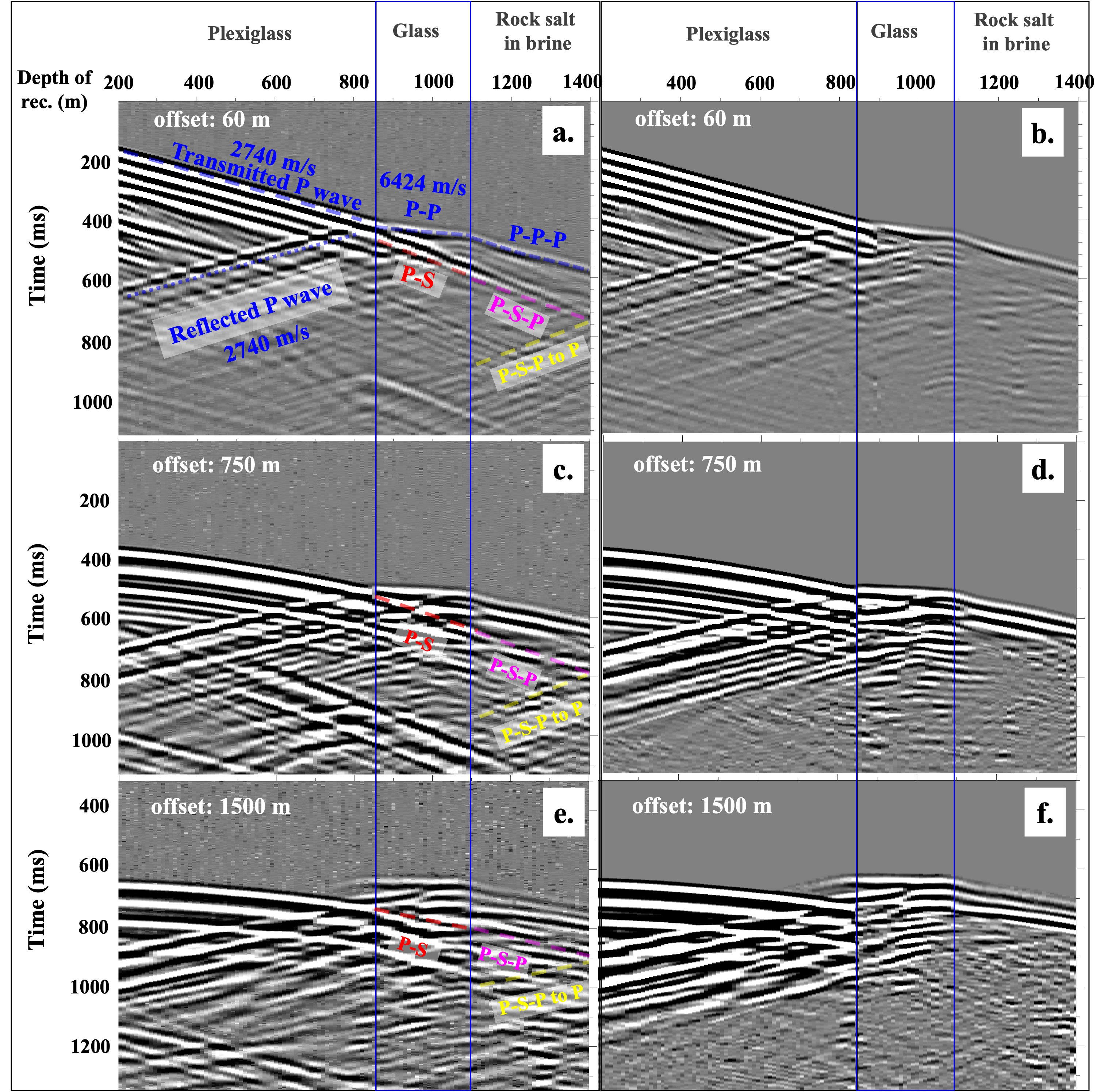}
\caption{The shot gathers for raw recordings at the offset locations of (a) 60 m, (c) 750 m, and (e) 1500 m are displayed. The corresponding separated pure P-mode wavefields are shown in (b), (d), and (f), respectively.}
\label{pm_shotgather}
\end{figure}

The raw seismic records for all three shots are displayed in Figure~\ref{pm_shotgather}a, c, and e, respectively. The boundaries for different layers are marked in their corresponding depths. The characteristics of wave propagation and conversion are observed to change with increasing depth and offset. In the near-offset shot, P-wave transmissions and reflections are indicated with dashed blue lines. In this experiment, a significant amount of transmission energy converts to shear mode (P-S mode, indicated by dashed red lines) at the top of the glass, where a sharp impedance increase occurs, even in the near-offset shot. The converted S-mode further propagates and converts to a P-S-P mode in brine before being received by the recording transducer. The energy of mode conversion increases with greater offset distance. To meet the prerequisites of the proposed 3D GPU-AWRI, the multi-mode P-wave and converted S-wave must be removed, leaving the pure P-wavefield, which includes both transmissions and reflections. Therefore, we proceed to separate the pure P-mode wavefields using a set of designated time-space domain and F-K domain filters, as shown in Figure~\ref{pm_shotgather}b, d, and f, for varying offset locations.

\subsection{Inversion Result}

The PM dataset contains only three sources, which is limited for directly inverting the entire 3D velocity profile. Given that the model is almost homogeneous laterally, we aim to invert the 1D velocity profile instead. In this example, we perform inversions using WRI and AWRI-TV with data at 7.5 Hz and 8 Hz. The model is discretized at 50 m intervals. We fine-tuned the penalty parameter to \(\lambda=1\times10^2\) and the TV weighting parameter to \(\gamma=1\times10^{-1}\). 

\begin{figure}
\centering
\subfloat[\label{fig:PMTrue}]{\includegraphics[width=0.500\hsize]{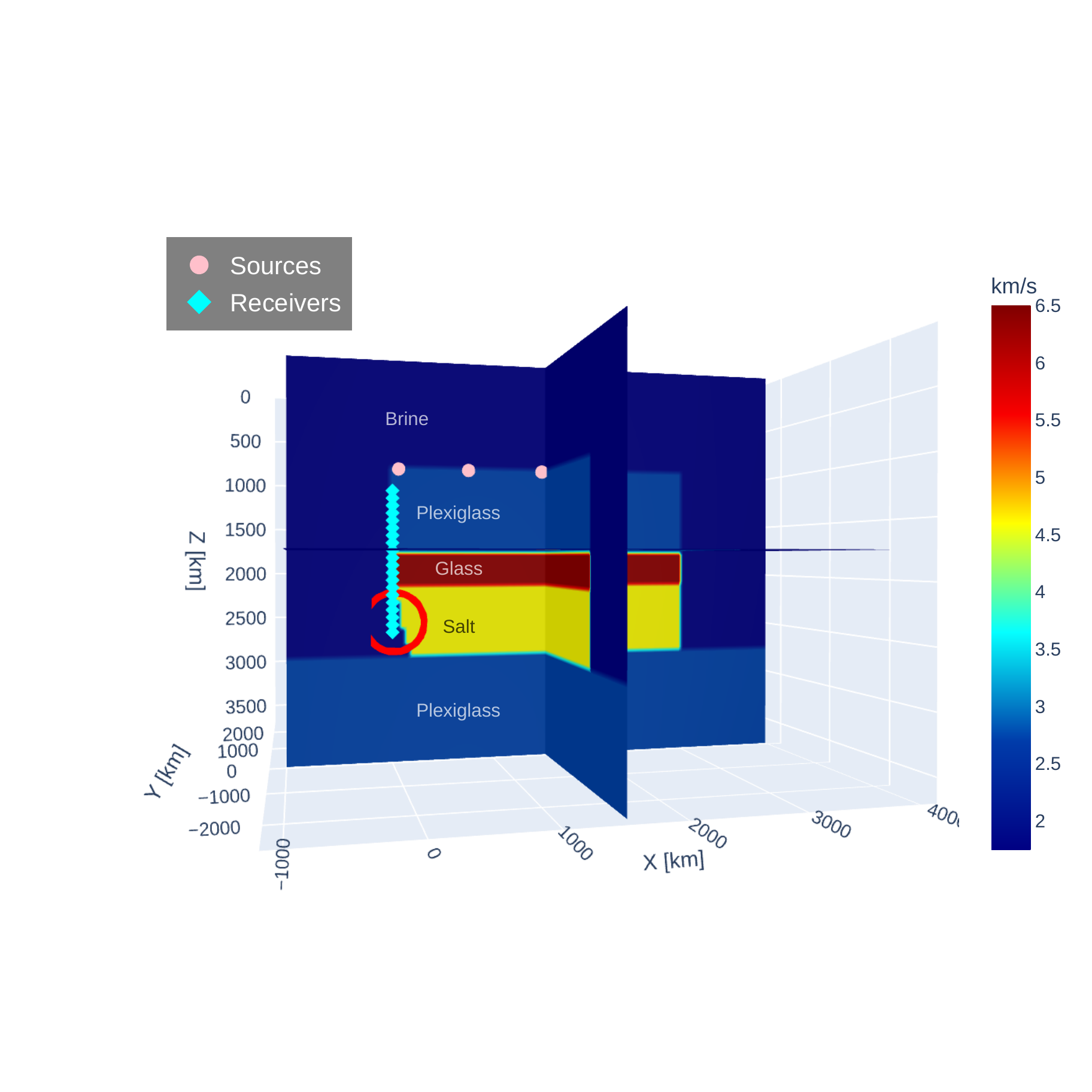}}\\
\subfloat[\label{fig:PMIni}]{\includegraphics[width=0.500\hsize]{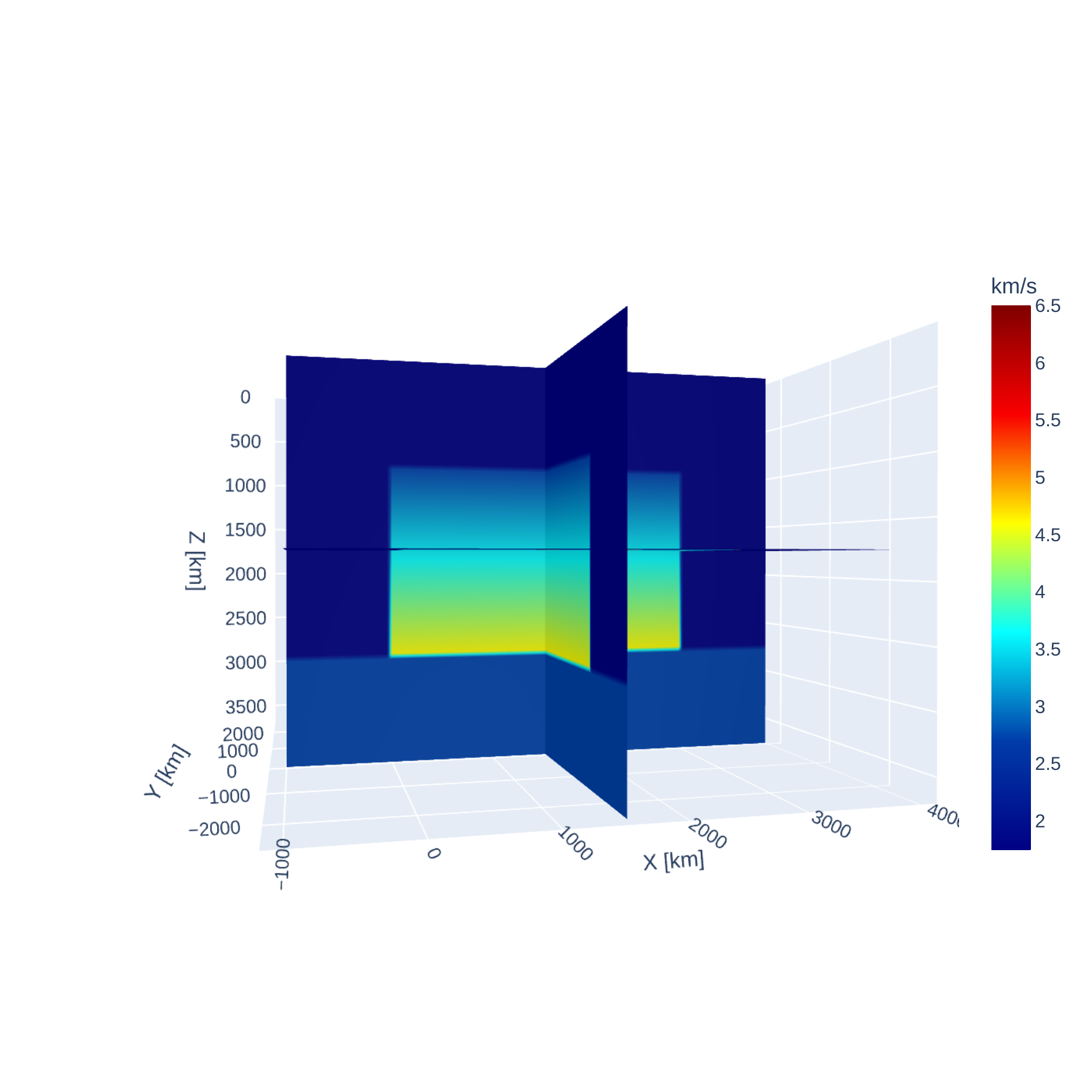}}
% \subfloat[\label{fig:MDEcmp}]{\includegraphics[width=0.400\hsize]{./Figure4a.eps}}
\caption{(a) True model and (b) initial model of the VSP PM example.}\label{fig:PModels}
\end{figure}

To conduct the inversion, we create a model depicted in Figure \ref{fig:PMTrue} to replicate the real experimental environment. The entire computational domain encompasses the sample, the surrounding brine, and the bottom of the water tank, which is made of Plexiglass. The total model size is 5 km \(\times\) 5 km \(\times\) 3.5 km, with the sample located at the center of the model. Source and receiver locations are indicated by solid circles and diamonds, respectively. During the inversion, we assume the velocities of the brine and Plexiglass are known and exclude them from inversion. Our goal is to invert a 1D velocity profile for the sample. The inversion is performed using the initial model shown in Figure \ref{fig:PMIni}.

\begin{figure}
\centering
\subfloat[\label{fig:PMWRI}]{\includegraphics[width=0.500\hsize]{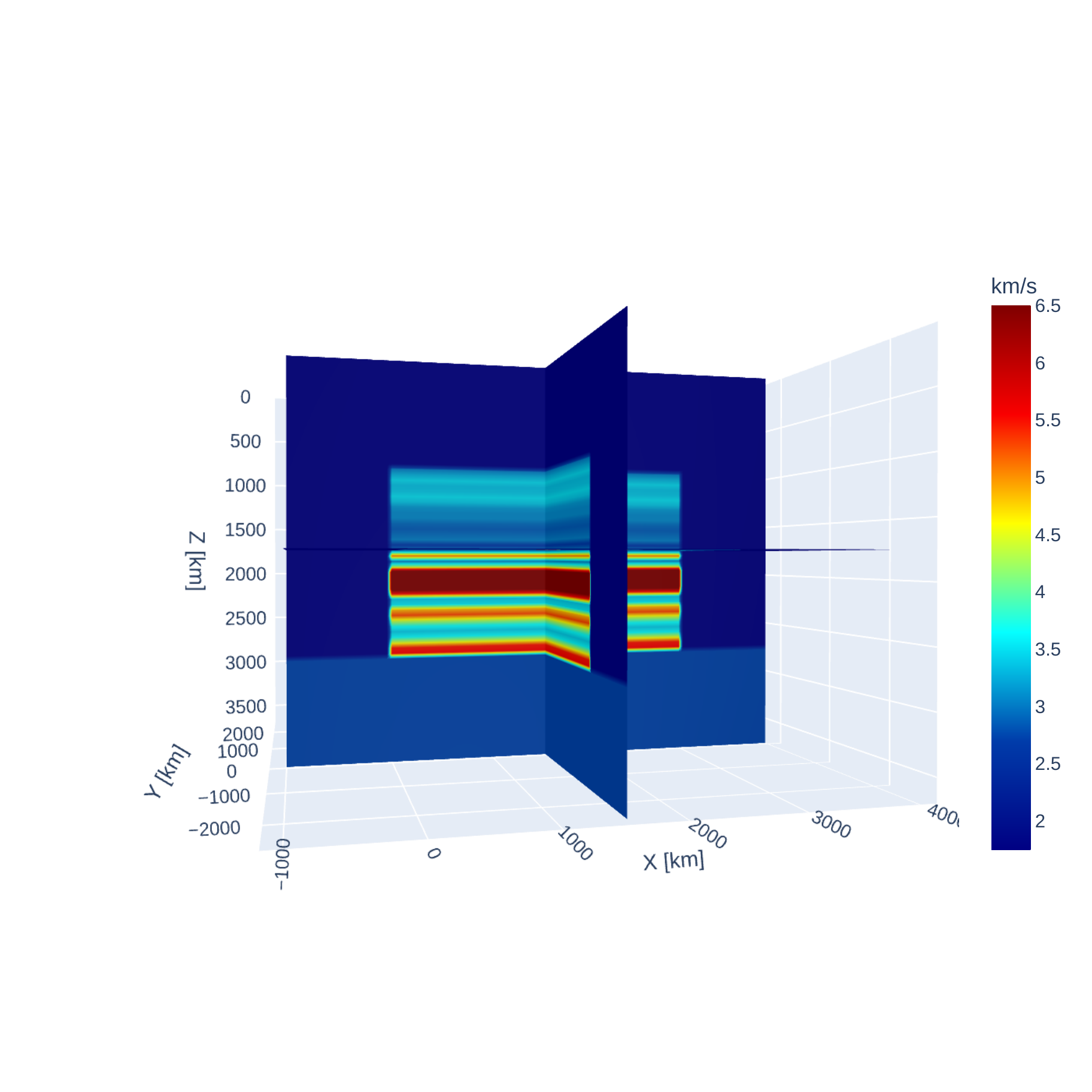}}\\
\subfloat[\label{fig:PMAWRI}]{\includegraphics[width=0.500\hsize]{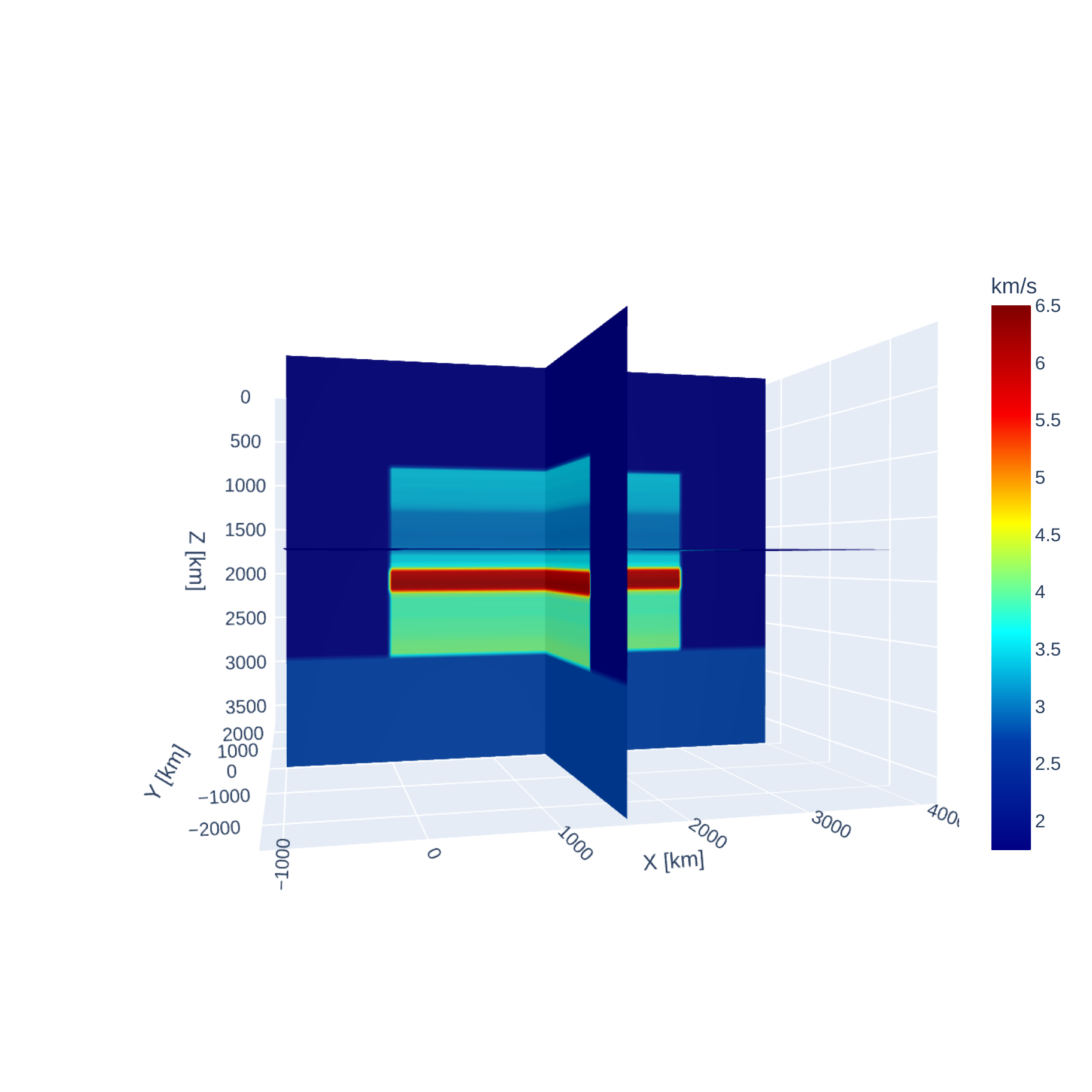}}
% \subfloat[\label{fig:MDEcmp}]{\includegraphics[width=0.400\hsize]{./Figure4a.eps}}
\caption{Inversion results of (a) WRI and (b) AWRI for the VSP PM example. }\label{fig:PMresults}
\end{figure}

\begin{figure}[!htbp]
\centering
\includegraphics[width=\hsize]{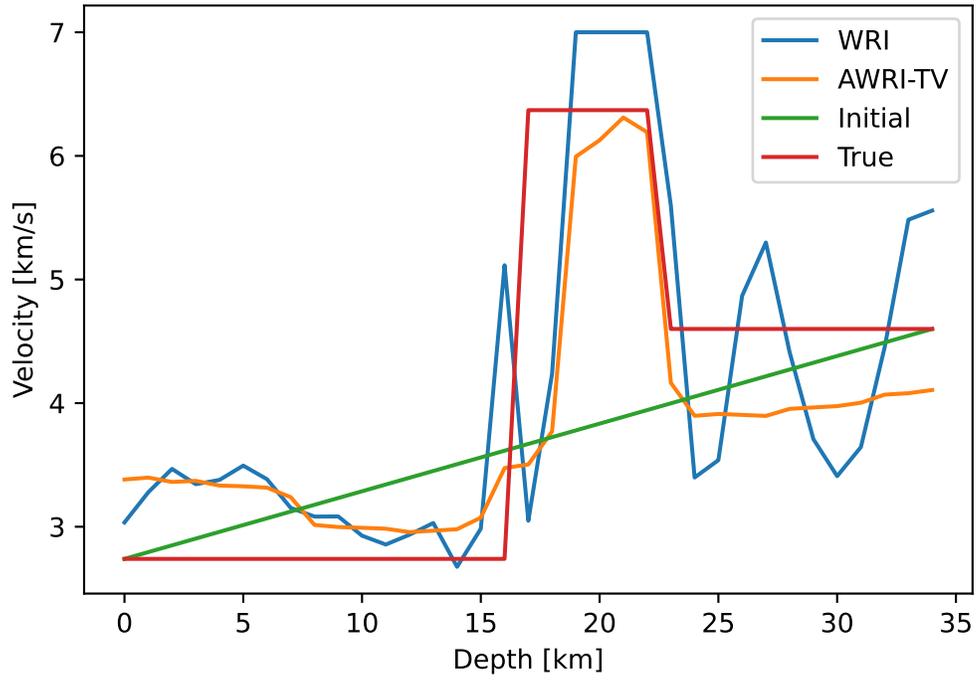}
\caption{Comparison of 1D Velocity Profiles at \(x = 500\) m and \(y = 0\) m: True Model vs. Initial Model vs. WRI and AWRI-TV Results.}
\label{fig:PMresultCmp}
\end{figure}

Figure \ref{fig:PMresults} presents the inversion results for WRI and AWRI-TV. Figure \ref{fig:PMresultCmp} compares the 1D velocity profiles at \(x = 500\) m and \(y = 0\) m among the true model, the initial model, and the results obtained by WRI and AWRI-TV. Table \ref{tab:PMResCmp} compares the relative model errors among the initial model, WRI results, and AWRI-TV results, along with the computational times for WRI and AWRI-TV. Both methods accurately reconstruct the high-velocity second layer. However, AWRI-TV produces more stable results than WRI due to the TV regularization, which smooths the model. In the third layer, WRI results exhibit oscillations, whereas AWRI-TV provides a more stable profile, albeit with a slightly lower velocity than the true model. This discrepancy occurs because the third layer features a tilted edge, originally designed to simulate a steeply-dipping salt flank. Consequently, the end portion of the vertical receiver array is immersed in super-saturated brine rather than being directly attached to the edge of the model, as shown in Figure \ref{fig:PMTrue}. Waves exit the salt body and travel through the brine before being picked up by the receiver, introducing challenges in inverting the 1D velocity profile. This is particularly problematic because the true model is not purely 1D, adding to the uncertainty. Given that brine has a much lower velocity than salt, the 1D velocity obtained by AWRI-TV is slightly lower than the true value, as illustrated in Figure \ref{fig:PMresultCmp}.

Despite uncertainties related to data acquisition, model geometry, and limited data, AWRI-TV effectively reconstructs the three-layer structure, demonstrating robustness in handling real data. Additionally, AWRI-TV achieves this using only 84\% of the computational time required by WRI, while delivering more stable results.

\begin{table}
    \centering
    \begin{tabular}{c|ccc}
        \hline
                              & Initial &WRI                                          & AWRI-TV  \\
        \hline
        Error($\%$)   & 28 & 25                                             & 24\\

         Time (s)       & -  &1932                                          & 1625\\
         \hline
    \end{tabular}
    \caption{Comparison of relative model errors and computational time for Initial model, WRI and AWRI-TV Results.}
    \label{tab:PMResCmp}
\end{table}

\section{Conclusion}\label{conclusion}

Wavefield reconstruction inversion has demonstrated potential as an approach to address the 'cycle-skipping' problem encountered by conventional FWI. However, its application to 3D problems with real data remains unexplored in the current research landscape. In this study, we introduce a 3D source-free adaptive WRI approach accelerated by GPU. Furthermore, we apply the proposed method to experimental VSP physical modeling data, marking, to the best of our knowledge, the inaugural reported real-data VSP application of a WRI-related method.

We first accelerate conventional WRI by implementing an adaptive accuracy control method. This adaptive selection of tolerance balances inversion accuracy and computational expenses. Additionally, incorporating TV regularization enhances the stability of AWRI at lower accuracy levels, further expediting the inversion process. As a result, the final AWRI-TV method not only achieves a 2-fold speed-up but also maintains inversion accuracy.

We further accelerate WRI by harnessing the parallel computing power of GPUs. Utilizing GPUs significantly speeds up computations in each LSQR iteration, resulting in a much faster WRI framework. Numerical examples demonstrate that with GPU assistance, we achieve a speed-up ratio of 195x (nearly 200x) compared to the CPU-based framework. Additional examples also show that the proposed 3D SF-AWRI can perform inversion effectively without requiring an accurate source signature function.

The numerical example for the VSP scenario indicates that the proposed GPU-SF-AWRI inherits WRI's ability to address local minima effectively. This demonstrates GPU-SF-AWRI’s robustness in avoiding convergence to suboptimal solutions, which is a common challenge in conventional FWI. Additionally, applying the proposed 3D GPU-SF-AWRI to real experimental VSP physical modeling data showcases the method’s practical applicability and effectiveness in dealing with authentic, complex datasets. This application confirms that GPU-SF-AWRI not only excels in synthetic examples but also performs reliably in real-world scenarios, handling intricacies such as noise and inaccuracies inherent in actual data. Equipped with TV regularization, the GPU-SF-AWRI not only reduces computational costs but also delivers more stable and accurate results compared to conventional WRI.

% \newpage
% \section{Acknowledge}\label{acknowledge}

% This study is supported by National Natural Science Foundation of China (Grant No. 41974150, 42174158), a Supporting Program for Outstanding Talent of the University of Electronic Science and Technology of China (No. 2019-QR-01), Project of Basic Scientific Research Operating Expenses of Central Universities ZYGX2019J071, ZYGX2020J013) , and Chengdu International Scientific Reserach Cooperation Project (2022-GH02-00049-HZ)

\bibliography{zfang}

\begin{thebibliography}{}
\itemsep0pt

\bibitem[Aghamiry et~al., 2019]{aghamiry2019admm}
Aghamiry, H.~S., A. Gholami, and S. Operto,  2019, Admm-based multiparameter
  wavefield reconstruction inversion in vti acoustic media with tv
  regularization: Geophysical Journal International, {\bfseries 219},
  1316--1333.

\bibitem[Aghamiry et~al., 2022]{aghamiry2022admm}
Aghamiry, H.~S., A. Gholami, S. Operto, and A. Malcolm,  2022, Admm-based
  full-waveform inversion for microseismic imaging: Geophysical Journal
  International, {\bfseries 228}, 259--274.

\bibitem[Barnier et~al., 2023]{barnier2023full}
Barnier, G., E. Biondi, R.~G. Clapp, and B. Biondi,  2023, Full-waveform
  inversion by model extension: Theory, design, and optimization: Geophysics,
  {\bfseries 88}, R579--R607.

\bibitem[Blias and Hughes, 2015]{blias20153d}
Blias, E., and B. Hughes,  2015, 3d vsp imaging. some general problems,
  {\itshape in} SEG Technical Program Expanded Abstracts 2015: Society of
  Exploration Geophysicists,  5630--5634.

\bibitem[Chi et~al., 2014]{chi2014full}
Chi, B., L. Dong, and Y. Liu,  2014, Full waveform inversion method using
  envelope objective function without low frequency data: Journal of Applied
  Geophysics, {\bfseries 109}, 36--46.

\bibitem[Engquist and Froese, 2014]{engquist2014application}
Engquist, B., and B.~D. Froese,  2014, Application of the {W}asserstein metric
  to seismic signals: Communications in Mathematical Sciences, {\bfseries 12},
  979--988.

\bibitem[Esser et~al., 2018]{esser2016tvr}
Esser, E., L. Guasch, T. van Leeuwen, A.~Y. Aravkin, and F.~J. Herrmann,  2018,
  Total-variation regularization strategies in full-waveform inversion: SIAM
  Journal on Imaging Sciences, {\bfseries 11}, 376--406; doi:
  \href{https://doi.org/10.1137/17M111328X}{10.1137/17M111328X}.

\bibitem[Fang et~al., 2018a]{fang2018uncertainty}
Fang, Z., C. Da~Silva, R. Kuske, and F.~J. Herrmann,  2018a, Uncertainty
  quantification for inverse problems with weak partial-differential-equation
  constraints: Geophysics, {\bfseries 83}, R629--R647.

\bibitem[Fang and Demanet, 2020]{fang2020lift}
Fang, Z., and L. Demanet,  2020, Lift and relax for pde-constrained inverse
  problems in seismic imaging: IEEE Transactions on Geoscience and Remote
  Sensing, {\bfseries 59}, 8034--8039.

\bibitem[Fang et~al., 2024]{fang2024source}
Fang, Z., H. Wang, M. Li, X. Gu, and P. Li,  2024, A source independent
  tv-regularized full waveform inversion method for peripheral imaging around a
  borehole: IEEE Transactions on Geoscience and Remote Sensing.

\bibitem[Fang et~al., 2018b]{fang2018source}
Fang, Z., R. Wang, and F.~J. Herrmann,  2018b, Source estimation for
  wavefield-reconstruction inversion: Geophysics, {\bfseries 83}, R345--R359.

\bibitem[Gholami et~al., 2022]{gholami2022extended}
Gholami, A., H. Aghamiry, and S. Operto,  2022, Extended full waveform
  inversion in the time domain by the augmented lagrangian method: Geophysics,
  {\bfseries 87}, R63--R77.

\bibitem[Golub and Pereyra, 2003]{golub2003separable}
Golub, G., and V. Pereyra,  2003, Separable nonlinear least squares: the
  variable projection method and its applications: Inverse Problems, {\bfseries
  19}, R1.

\bibitem[Huang et~al., 2017]{huang2017full}
Huang, G., R. Nammour, and W. Symes,  2017, Full-waveform inversion via
  source-receiver extension: Geophysics, {\bfseries 82(3)}, R153--R171.

\bibitem[Li et~al., 2014]{li2014wave}
Li, Y., B. Biondi, R. Clapp, and D. Nichols,  2014, Wave-equation migration
  velocity analysis for vti models: Geophysics, {\bfseries 79}, WA59--WA68.

\bibitem[Lin et~al., 2023]{lin2023fast}
Lin, Y., T. van Leeuwen, H. Liu, J. Sun, and L. Xing,  2023, A fast wavefield
  reconstruction inversion solution in the frequency domain: Geophysics,
  {\bfseries 88}, R257--R267.

\bibitem[M{\'e}tivier et~al., 2016]{metivier2016measuring}
M{\'e}tivier, L., R. Brossier, Q. M{\'e}rigot, E. Oudet, and J. Virieux,  2016,
  Measuring the misfit between seismograms using an optimal transport distance:
  application to full waveform inversion: Geophysical Supplements to the
  Monthly Notices of the Royal Astronomical Society, {\bfseries 205}, 345--377.

\bibitem[Nocedal and Wright, 2000]{Nocedal:2000}
Nocedal, J., and S.~J. Wright,  2000, {Numerical optimization}: Springer.

\bibitem[Operto et~al., 2023]{operto_extending_2023}
Operto, S., A. Gholami, H. Aghamiry, G. Guo, S. Beller, K. Aghazade, F.
  Mamfoumbi, L. Combe, and A. Ribodetti,  2023, Extending the search space of
  full-waveform inversion beyond the single-scattering {Born} approximation:
  {A} tutorial review: Geophysics, {\bfseries 88}, R671--R702; doi:
  \href{https://doi.org/10.1190/geo2022-0758.1}{10.1190/geo2022-0758.1}.

\bibitem[Paige and Saunders, 1982]{paige1982lsqr}
Paige, C.~C., and M.~A. Saunders,  1982, Lsqr: An algorithm for sparse linear
  equations and sparse least squares: ACM Transactions on Mathematical Software
  (TOMS), {\bfseries 8}, 43--71.

\bibitem[Pratt et~al., 1998]{Pratt98}
Pratt, G., C. Shin, and G. Hicks,  1998, {Gauss-Newton and full Newton methods
  in frequency-space seismic waveform inversion}: Geophysical Journal
  International, {\bfseries 133}, 341--362; doi:
  \href{https://doi.org/10.1046/j.1365-246X.1998.00498.x}{10.1046/j.1365-246X.1998.00498.x}.

\bibitem[Rizzuti et~al., 2021]{rizzuti2021dual}
Rizzuti, G., M. Louboutin, R. Wang, and F.~J. Herrmann,  2021, A dual
  formulation of wavefield reconstruction inversion for large-scale seismic
  inversion: Geophysics, {\bfseries 86}, R879--R893.

\bibitem[Song et~al., 2023]{song2023elastic}
Song, C., X. Feng, Y. Gao, B. Li, and C. Liu,  2023, Elastic wavefield
  reconstruction inversion with source estimation: IEEE Transactions on
  Geoscience and Remote Sensing, {\bfseries 62}, 1--9.

\bibitem[Stewart et~al., 2002]{stewart2002converted}
Stewart, R.~R., J.~E. Gaiser, R.~J. Brown, and D.~C. Lawton,  2002,
  Converted-wave seismic exploration: {M}ethods: Geophysics, {\bfseries 67},
  1348--1363.

\bibitem[Symes, 2008]{symes2008migration}
Symes, W.~W.,  2008, Migration velocity analysis and waveform inversion:
  Geophysical prospecting, {\bfseries 56}, 765--790.

\bibitem[Tarantola and Valette, 1982]{Tarantola1982FWI}
Tarantola, A., and B. Valette,  1982, Generalized nonlinear inverse problems
  solved using the least squares criterion: Reviews of Geophysics, {\bfseries
  20}, 219--232; doi:
  \href{https://doi.org/10.1029/RG020i002p00219}{10.1029/RG020i002p00219}.

\bibitem[Tromp, 2019]{tromp2019seismic}
Tromp, J.,  2019, Seismic wavefield imaging of {E}arth's interior across
  scales: Nature Reviews Earth \& Environment,  1--14.

\bibitem[{van Leeuwen} and Herrmann, 2013]{vanLeeuwen2013GJImlm}
{van Leeuwen}, T., and F.~J. Herrmann,  {2013}, Mitigating local minima in
  full-waveform inversion by expanding the search space: Geophysical Journal
  International, {\bfseries 195}, 661--667; doi:
  \href{https://doi.org/10.1093/gji/ggt258}{10.1093/gji/ggt258}.

\bibitem[van Leeuwen and Herrmann, 2014]{van20143d}
--------, 2014, 3d frequency-domain seismic inversion with controlled
  sloppiness: SIAM Journal on Scientific Computing, {\bfseries 36}, S192--S217.

\bibitem[{van Leeuwen} and Herrmann, 2015]{vanleeuwen2015IPpmp}
--------, 2015, A penalty method for {PDE}-constrained optimization in inverse
  problems: Inverse Problems, {\bfseries 32}, 015007.

\bibitem[{van Leeuwen} and Mulder, 2010]{van2010correlation}
{van Leeuwen}, T., and W. Mulder,  2010, A correlation-based misfit criterion
  for wave-equation traveltime tomography: Geophysical Journal International,
  {\bfseries 182}, 1383--1394.

\bibitem[Virieux and Operto, 2009]{VirieuxOverview2009}
Virieux, J., and S. Operto,  2009, An overview of full-waveform inversion in
  exploration geophysics: Geophysics, {\bfseries 74(6)}, WCC1--WCC26; doi:
  \href{https://doi.org/10.1190/1.3238367}{10.1190/1.3238367}.

\bibitem[Warner and Guasch, 2016]{warner2016adaptive}
Warner, M., and L. Guasch,  2016, Adaptive waveform inversion: Theory:
  Geophysics, {\bfseries 81(6)}, R429--R445.

\bibitem[Warner et~al., 2013]{warner2013full}
Warner, M., T. Nangoo, N. Shah, A. Umpleby, J. Morgan, et~al.,  2013,
  Full-waveform inversion of cycle-skipped seismic data by frequency
  down-shifting: 83th Annual International Meeting, SEG, Expanded Abstracts,
  903--907.

\bibitem[Wu et~al., 2013]{wu2013ultra}
Wu, R.-S., J. Luo, and B. Wu,  2013, Ultra-low-frequency information in seismic
  data and envelope inversion, {\itshape in} SEG Technical Program Expanded
  Abstracts 2013: Society of Exploration Geophysicists,  3078--3082.

\bibitem[Yang et~al., 2020]{yang2020least}
Yang, J., Y.~E. Li, Y. Liu, and J. Zong,  2020, Least-squares extended reverse
  time migration with randomly sampled space shifts: Geophysics, {\bfseries
  85}, S357--S369.

\bibitem[Yang et~al., 2018]{yang2018application}
Yang, Y., B. Engquist, J. Sun, and B.~F. Hamfeldt,  2018, Application of
  optimal transport and the quadratic wasserstein metric to full-waveform
  inversion: Geophysics, {\bfseries 83(1)}, R43--R62.

\end{thebibliography}

\end{document}